\documentclass[oldversion]{aa}
\usepackage{graphicx}
\usepackage{txfonts}
\usepackage{natbib}
\usepackage{longtable}

\def\ltsima{$\; \buildrel < \over \sim \;$}
\def\simlt{\lower.5ex\hbox{\ltsima}}
\def\gtsima{$\; \buildrel > \over \sim \;$}
\def\simgt{\lower.5ex\hbox{\gtsima}}

\newcommand \beq {\begin{equation}}
\newcommand \eeq {\end{equation}}

\def\npar{\hbox{\hspace{1.0mm}}
       \vbox{\hbox{$\parallel$}}
       \vbox{\moveleft6.3pt\hbox{$\not$}}\hbox{\hskip 0.5mm}}

\begin{document}
\title{Swift observations of GRB~060614: an anomalous burst with a well behaved afterglow}
\author{
V.~Mangano\inst{1} \and 
S.~T.~Holland\inst{2,3} \and 
D.~Malesani\inst{4} \and 
E.~Troja\inst{1,5,6} \and 
G.~Chincarini\inst{7,8} \and 
B.~Zhang\inst{9} \and 
V.~La~Parola\inst{1,7} \and 
P.~J.~Brown\inst{10} \and 
D.~N.~Burrows\inst{10} \and 
S.~Campana\inst{8} \and 
M.~Capalbi\inst{11} \and 
G.~Cusumano\inst{1} \and 
M.~Della Valle\inst{12} \and 
N.~Gehrels\inst{2} \and 
P.~Giommi\inst{11} \and 
D.~Grupe\inst{10} \and 
C.~Guidorzi\inst{7,8} \and 
T.~Mineo\inst{1} \and 
A.~Moretti\inst{8} \and 
J.~P.~Osborne\inst{5} \and 
S.~B.~Pandey\inst{13} \and 
M.~Perri\inst{11} \and 
P.~Romano\inst{7,8} \and 
P.~W.~A.~Roming\inst{10} \and 
G.~Tagliaferri\inst{8}
}

\offprints{V.~Mangano, INAF--Istituto di Astrofisica Spaziale 
                 e Fisica Cosmica Sezione di Palermo, 
                 via Ugo La Malfa 153, I-90146 Palermo, Italy; 
                 \email{vanessa@ifc.inaf.it}}

\institute{INAF -- Istituto di Astrofisica Spaziale 
                 e Fisica Cosmica Sezione di Palermo, 
                 via Ugo La Malfa 153, I-90146 Palermo, Italy
            \and NASA/Goddard Space Flight Center, Greenbelt, MD 20771, USA
            \and Universities Space Research Association, 10227 Wincopin Circle,
                 Suite 221, Columbia, MD 21044 USA
            \and Dark Cosmology Centre, Niels Bohr Institut, University of Copenhagen,
                 Juliane Maries vej 30, DK--2100 K\o{}benhavn, Denmark 
            \and Department of Physics and Astronomy, University of Leicester,
                 Leicester LE1 7RH, UK
            \and Dipartimento di Scienze Fisiche ed Astronomiche, Sezione di
                 Astronomia, Universit\`a di Palermo, piazza del Parlamento 1, I-90134
                 Palermo, Italy
            \and Universit\`a degli studi di Milano-Bicocca,
                 Dipartimento di Fisica, piazza delle Scienze 3, 
                 I-20126 Milano, Italy
            \and INAF -- Osservatorio Astronomico di Brera, 
                 via Emilio Bianchi 46, I-23807 Merate (LC), Italy
            \and Department of Physics, University of Nevada, Las Vegas, NV 89154, USA
            \and Department of Astronomy \& Astrophysics, 525 Davey Lab., 
                 Pennsylvania State University, University Park, PA 16802, USA
            \and ASI Science Data Center, via Galileo Galilei, 
                 I-00044 Frascati (Roma), Italy
            \and INAF -- Osservatorio Astrofisico di Arcetri, largo E. Fermi 5, 
                 I-50125 Firenze, Italy 
            \and Mullard Space Science Laboratory, University College of London, 
                 Holmbury St Mary, Dorking, Surrey, RH5 6NT, UK
                 }

\date{Received: - /  Accepted -}
\abstract{
GRB~060614 is a remarkable gamma-ray burst (GRB) observed by Swift
with puzzling properties, which challenge current progenitor models.
In particular, the lack of any bright supernova (SN) down to
very strict limits and the vanishing spectral lags 
{\rm during the whole burst} are typical of short GRBs, strikingly at odds 
with the long (102~s) duration of this event.
Here we present detailed spectral and temporal analysis of the Swift
observations  of GRB~060614. 
We show that the burst presents standard optical, ultraviolet and X-ray afterglows, 
detected beginning 4 ks after the trigger. An achromatic break is observed
simultaneously in the optical and X-ray bands, at a time consistent with
the break in the $R$-band light curve measured by the VLT. The achromatic
behaviour and the consistent post-break decay slopes make GRB\,060614 one
of the best examples of a jet break for a Swift burst.
The optical and ultraviolet afterglow light curves have also 
an earlier break at 29.7$\pm$4.4~ks, marginally consistent 
with a corresponding break at 36.6$\pm$2.4~ks observed in the X-rays. 
In the optical, there is strong spectral evolution around this break, suggesting the
passage of a break frequency through the optical/ultraviolet band. 
The very blue spectrum at early times suggests this may be the injection
frequency, as also supported by the trend in the light curves: rising at
low frequencies, and decaying at higher energies.
The early X-ray light curve (from 97 to 480 s) is well interpreted as the X-ray 
counterpart of the burst extended emission. 
Spectral analysis of the BAT and XRT data in the $\sim 80$ s overlap time interval 
show that the peak energy of the burst has decreased to as low as 8~keV at the beginning 
of the XRT observation.
Spectral analysis of following XRT data shows that the peak energy of the burst
continues to decrease through the XRT energy band and exits it at about 500~s 
after the trigger.
The average peak energy $E_{\rm p}$ of the burst is likely below 
the BAT energy band ($< 24$~keV at the 90\% confidence level) but larger than 8 keV.
{\rm The initial group of peaks observed by BAT ($\sim$ 5 s) is however
distinctly harder than the rest of the prompt emission, with a peak energy  of
about 300 keV as measured by Konus Wind. Considering the time-averaged spectral
properties, GRB~060614 is consistent} with the $E_{\rm iso}-E_{\rm p}^{\rm
rest}$, $E_{\gamma}-E_{\rm p}^{\rm rest}$, and $L_{\rm p,iso}-E_{\rm p}^{\rm
rest}$ correlations. 
\keywords{gamma rays: bursts; X-rays: individual (GRB~060614)}
}

\authorrunning{ V.~Mangano et al.}
\titlerunning{Swift observations of GRB~060614}
\maketitle

\section{Introduction}
\label{section:introduction}

The Swift Gamma-Ray Burst Explorer (\citealt{gehr}), 
successfully launched on 2004 Nov.\ 20, is a multi-wavelength
space observatory with a payload that includes one wide-field instrument, 
the Burst Alert Telescope (BAT, 15$-$350 keV energy band; \citealt{barth05a}),  
and two narrow-field instruments (NFIs), 
the X-Ray Telescope (XRT, 0.2$-$10\,keV; \citealt{bur05a}) 
and the Ultraviolet/Optical Telescope (UVOT, 1700$-$6500\,\AA; \citealt{rom05}).
BAT has been designed for burst detection and localization to $\la 3\arcmin$ accuracy.
It triggers an autonomous slew of the observatory to point the two narrow-field instruments,
typically within 100\,s from the burst onset.
XRT can provide $\la 5\arcsec$ positions, 
while UVOT further refines the afterglow localization to $\sim 0\farcs5$. 

Swift-BAT triggered on GRB~060614 (trigger~214805) 
on 2006 June 14 at 12:43:48 UT, and located it 
at the coordinates RA$_{\rm J2000} = 21 ^{\rm h} 23 ^{\rm m} 27^{\rm s}$, 
Dec$_{\rm J2000} = -53\degr 02\arcmin 02\arcsec$,
with an uncertainty of 3$\arcmin$ (90\% containment, including 
systematic uncertainty; \citealt{gcn5252}).
The spacecraft executed an automatic and immediate slew 
to the burst location and started observing with XRT and UVOT
91 and 101~s after the trigger, respectively.
The afterglow emission was monitored for more than 30 days.

The BAT mask-weighted light curve showed a multi-peaked structure beginning
with an initial {\rm $\sim$5~s series of hard, bright  peaks} followed by a
fainter,  softer and {\rm highly variable} extended prompt emission. The
observed fluence in the 15$-$150 keV energy band was estimated at the level of
(2.17$\pm$0.04)$\times$10$^{-5}$~erg~cm$^{-2}$ for a burst duration of
$T_{90}=\,102~\pm~5$~s  \citep[15$-$350~keV;][]{gcn5256}.  This is one of the
highest fluences ever observed for a Swift bursts located by BAT.

XRT found a very bright ($\sim$1300\,\,counts~s$^{-1}$) 
uncatalogued source inside the BAT error circle. The object 
showed an initial fast exponential decay and then a flattening
at the level of $\sim$0.2\,counts~s$^{-1}$, followed by a steepening
to a standard afterglow evolution. 
The initial decay was accompanied by strong hard-to-soft 
spectral evolution \citep{gcn5254}.

The optical counterpart, originally detected in the White (160$-$650 nm) 
filter at the level of 18.4$\pm$0.5 mag, was later visible in all 
of the UVOT bands (White, $V$, $B$, $U$, UVW1, UVM2, and UVW2 filters).
The detection in the UVW2 filter allows us to set a strong upper
limit to the burst redshift ($z<1.3$ at the 99.99\% confidence level;
\citealp{gehrels06}) and implies very low dust extinction.

Konus-Wind was also triggered by GRB~060614 at 12:43:51.59 UT, 
$\sim$4~s after the BAT trigger.
The spectrum of the first {\rm group of intense peaks} was {\rm fitted} 
in the 20~keV$-$2~MeV energy range by a power law with an exponential 
cut-off model ($\chi^2/\mbox{d.o.f.}=73/59$). 
The derived photon index was 1.57$^{+0.12}_{-0.14}$ with 
a peak energy $E_{\rm p} = 302^{+214}_{-85}$~keV.
The spectrum of the remaining part of the prompt emission was 
described by a simple power law model with photon index 2.13$\pm$0.05.
The total fluence in the 20~keV$-$2~MeV energy band was 
4.09$^{+0.18}_{-0.34}$ $\times$ 10$^{-5}$~erg~cm$^{-2}$ \citep{gcn5264}.

Ground-based optical and infrared follow-up observations were performed with
several instruments.
Optical/infrared imaging, made with the ANDICAM instrument on the 1.3~m telescope at the
Cerro Tololo Inter-American Observatory (CTIO),
revealed the GRB~060614 afterglow $\sim$15.5~hr after the trigger both in the $I$ and $J$ bands
\citep{gcn5259,cobb06}. 
In the $R$ band, the afterglow was first detected by the Siding Spring
Observatory (SSO) 1~m telescope, brightening from a magnitude of $R=20.2\pm0.3$
$\sim$25~min  after the BAT trigger to $R=18.8\pm0.1$ after $\sim$6~hr
\citep{gcn5258}. 
Later observations were performed by the Watcher 0.4~m Telescope 
located in Boyden Observatory, South Africa
($R=19.0\pm0.3$, 7.1~hr after the trigger; \citealt{gcn5257}),
by the ESO VLT-UT1 ($R=19.3\pm0.2$ after $\sim$14.4~hr;
\citealt{gcn5261,dellavalle06}) 
and by the Danish 1.5~m Telescope equipped with the Danish Faint Object 
Spectrograph and Camera (DFOSC; \citealt{gcn5272,fynbo06}). 
The last two monitored the source for several weeks.
The observed flattening of the optical and infrared emission few days after 
the burst was attributed to the host galaxy by \citet{gcn5277} and \citet{gcn5282}.

Based on the detection of the host galaxy emission lines, a redshift of 
$z=0.125$ was proposed by \citet{gcn5275} and confirmed by \citet{gcn5276}.
The GRB host is a faint ($M_{V}=-15.5$) star-forming galaxy \citep{dellavalle06,fynbo06}
with a specific star formation rate at the low end of the distribution for long GRB hosts 
\citep{christensen04,sollerman05}. 
The GRB counterpart is located in the outskirts of the host \citep{galyam06}. 
The probability of chance alignment between the GRB and this galaxy
has been carefully estimated to be as low as $5.6\times10^{-6}$ by 
\citet{galyam06}.

The long monitoring campaigns of the ESO VLT and of the Danish 1.5~m Telescope 
at La Silla in Chile, and the target of opportunity observations
of the Hubble Space Telescope (HST)
did not detect any SN component emerging out of the host galaxy light
\citep{dellavalle06,fynbo06,galyam06}. 
Any associated SN had to be more than 100 times fainter than events previously known
to be associated with long GRBs
(SN1998bw/GRB980425, \citealt{galama98}; 
SN2003dh/GRB030329, \citealt{stanek03,hjorth03};
SN2003lw/GRB031203, \citealt{malesani04};
SN2006aj/GRB~060218, \citealt{campana06}).
These robust limits might suggest either that GRB~060614 has
been produced {\rm during a merger process, or during the explosion 
of a ``fall back'' SN \citep{nomoto04}, or by a ``dark hypernova''
\citep{nomoto07}. Recently, \citet{tominaga07} showed through 
numerical simulations that jet-induced explosions in metal-poor 
massive stars can produce faint type-II SNe or dark HNe consistent
with existing upper limits.}  

\citet{gehrels06} noted that the first 5~s 
{\rm of the prompt emission, including the brightest peaks} 
of the BAT light curve, show many sub-pulses with 
time lags consistent with zero, like short bursts
and unlike long GRBs, which usually have positive lags \citep{norris02}.
In particular, in the peak luminosity$-$time lag plane, 
{\rm the brightest peak of} GRB~060614 lies in the
region occupied by short bursts \citep{gehrels06}.
Another similarity with short bursts is the structure of the BAT
light curve, starting with a {\rm series of bright, hard peaks}, 
followed by a {\rm group of lower luminosity, softer peaks and a smooth} tail. 
This is reminiscent of the long soft hump seen in several short GRBs
such as  GRB~050709 \citep{villasenor05}, GRB~050724 \citep{barth05b,campana06b},
GRB~051227 \citep{gcn4401}, GRB~060121 (HETE-2; \citealp{donaghylamb06}), 
GRB~061006 \citep{gcn5704}, GRB~061210 \citep{gcn5905},
and a few bursts in the BATSE sample (\citealt{norrisbonnel06}; see also
\citealt{lazzati01}).
Indeed \citet{zhang07} showed that GRB~060614, were it 8 times less energetics,
would have been detected by BATSE as a marginal short-duration GRB and by Swift
as an analog to GRB~050724.

However, {\rm current models of compact binary merger progenitors}
can hardly account for $\sim$100~s prolonged emission \citep{rosswog03,lee04},
{\rm even if the complexity of the physics involved does not allow
to set firm conclusions yet.}
In addition we note that recently \citet{amati06} have shown that
GRB~060614 is consistent with the $E_{\rm p}$ vs. $E_{\rm iso}$ relationship 
which applies only to long-duration GRBs \citep{newamati}. 
For these reasons GRB~060614 has been proposed to belong to a new
class of GRBs sharing observational properties with both the
long and short GRBs \citep{gehrels06} 
and possibly coming from different progenitors \citep{king07}.

Here we present a detailed analysis of Swift observations of GRB~060614. 
Details on the BAT, XRT and UVOT observations are given
in \S~\ref{section:data}; data reduction is described in 
\S~\ref{section:batanalysis}, \ref{section:xrtanalysis}, and 
\ref{section:uvotanalysis}; 
the temporal and spectral analysis results are reported in 
\S~\ref{section:analysis}, 
a summary of the results and discussion are presented in 
\S~\ref{section:results}.
Conclusions are drawn in \S~\ref{section:conclusions}. 

Throughout this paper the quoted uncertainties are given at 90\% 
confidence level for one interesting parameter (i.e., $\Delta \chi^2 =2.71$) 
unless otherwise stated.
Times are referred to the BAT trigger $T_0$, $t=T-T_0$, unless otherwise specified. 
We also adopt the notation $F_{\nu}(\nu,t) \propto t^{-\alpha} \nu^{-\beta}$
for the afterglow monochromatic flux 
as a function of time, where $\nu$ represents the frequency of the observed 
radiation. The energy index $\beta$ is related to the photon index 
$\Gamma$ according to $\beta = \Gamma-1$.
We adopt a standard cosmology model with $H_0 = 70$ km s$^{-1}$ Mpc$^{-1}$, 
$\Omega_{\rm M} = 0.3$, $\Omega_\Lambda = 0.7$.


\setcounter{table}{2}
 \begin{table}[ht]
 \begin{center}         
 \caption{Results of BAT time resolved spectral analysis.}     
 \label{batspecfit}
 \normalsize
 \begin{tabular}{llllll} 
 \hline 
 \hline 
 \noalign{\smallskip} 
 Spectrum  &  Start        &  End       &  $\Gamma$   &   d.o.f. & $\chi^2_{\rm r}$\\
           &  (s)          &  (s)       &             &          &           \\
 \noalign{\smallskip} 
 \hline 
 \noalign{\smallskip} 

BAT      & ~-2.83  &  176.5       & 2.13$_{-0.04 }^{+0.04 }$  &  56  &   0.73   \\
                                                                              
  \noalign{\smallskip}                                                        
\hline                                                                        
  \noalign{\smallskip}                                                        
                                                                              
BAT-A    & ~-2.83  &  ~~5.62      & 1.63$_{-0.07 }^{+0.07 }$  &  56  &   0.86   \\
 \noalign{\smallskip}                                                         
BAT-B    & ~~5.62  &  ~97.0       & 2.21$_{-0.04 }^{+0.04 }$  &  56  &   0.73   \\
 \noalign{\smallskip}                                                         
BAT-C    & ~97.0   &  176.5       & 2.37$_{-0.13 }^{+0.13 }$  &  56  &   0.76   \\
                                                                              
  \noalign{\smallskip}                                                        
\hline                                                                        
  \noalign{\smallskip}                                                        
                                                                              
BAT-1    & ~-2.83  &   ~-1.0      & 1.51$_{-0.19 }^{+0.20 }$  &  56  &   0.93   \\    
 \noalign{\smallskip}                                                         
BAT-2    & ~-1.0   &   ~~1.0      & 1.56$_{-0.09 }^{+0.09 }$  &  56  &    1.31  \\
 \noalign{\smallskip}                                                         
BAT-3    & ~~1.0   &   ~~4.0      & 1.78$_{-0.10 }^{+0.10 }$  &  56  &   0.75   \\
 \noalign{\smallskip}                                                         
BAT-4    & ~~4.0   &   ~10.0      & 1.92$_{-0.24 }^{+0.25 }$  &  56  &   0.83   \\
 \noalign{\smallskip}                                                         
BAT-5    & ~10.0   &   ~17.0      & 1.84$_{-0.13 }^{+0.13 }$  &  56  &   0.85   \\
 \noalign{\smallskip}                                                         
BAT-6    & ~17.0   &   ~23.0      & 2.05$_{-0.10 }^{+0.11 }$  &  56  &    1.01  \\
 \noalign{\smallskip}                                                         
BAT-7    & ~23.0   &   ~30.0      & 2.14$_{-0.11 }^{+0.11 }$  &  56  &   0.71   \\
 \noalign{\smallskip}                                                         
BAT-8    & ~30.0   &   ~33.0      & 2.07$_{-0.10 }^{+0.10 }$  &  56  &   0.62   \\
 \noalign{\smallskip}                                                         
BAT-9    & ~33.0   &   ~37.8      & 2.33$_{-0.10 }^{+0.10 }$  &  56  &   0.88   \\
 \noalign{\smallskip}                                                         
BAT-10   & ~37.8   &   ~42.5      & 2.17$_{-0.07 }^{+0.07 }$  &  56  &    1.02  \\
 \noalign{\smallskip}                                                         
BAT-11   & ~42.5   &   ~47.5      & 2.22$_{-0.06 }^{+0.06 }$  &  56  &   0.92   \\
 \noalign{\smallskip}                                                         
BAT-12   & ~47.5   &   ~60.0      & 2.07$_{-0.05 }^{+0.05 }$  &  56  &    1.03  \\
 \noalign{\smallskip}                                                         
BAT-13   & ~60.0   &   ~80.0      & 2.24$_{-0.06 }^{+0.06 }$  &  56  &   0.79   \\
 \noalign{\smallskip}                                                         
BAT-14   & ~80.0   &   ~97.0      & 2.14$_{-0.08 }^{+0.08 }$  &  56  &   0.75   \\
 \noalign{\smallskip}                                                         
BAT-15   & ~97.0   &   128.0      & 2.46$_{-0.12 }^{+0.12 }$  &  56  &    1.09  \\
 \noalign{\smallskip}                                                         
BAT-16   & 128.0   &   175.0      & 2.19$_{-0.21 }^{+0.22 }$  &  56  &   0.66   \\

  \noalign{\smallskip}
  \hline
  \end{tabular}
  \end{center}
  \end{table} 

\section{Observations and data reduction}
\label{section:data}

The list of all BAT, XRT and UVOT observations
of GRB~060614 used for the present analysis
is shown in Tables~\ref{log} and \ref{loguvot}. 
XRT and UVOT follow-up lasted 51 days and consisted
of 38 sequences numbered from 0 to 37.
The total XRT exposure spent on GRB~060614 was 514 ks.
For both XRT and UVOT, merging data of many sequences 
was necessary to attain source detection at later times.
In these cases only information about the merged data is
reported.


\subsection{BAT}
\label{section:batanalysis}

\begin{figure}[hb]
\includegraphics[width=5.3cm,angle=-90]{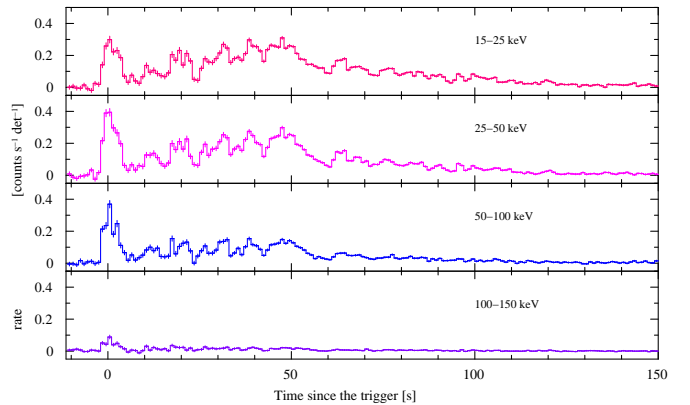}
\caption{BAT light curves in four energy bands.}
\label{batlc}
\end{figure}

\begin{figure}[hb]
\includegraphics[width=6.8cm,angle=-90]{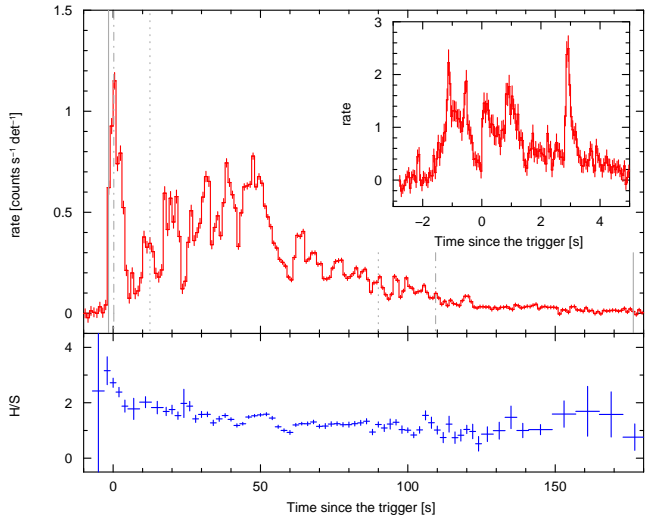}
\caption{In the upper panel the 15$-$150~keV BAT light curve is shown at 1~s
time resolution.  Solid (dotted-dashed) vertical lines mark the $T_{100}$
($T_{90}$) time interval.  Dotted vertical lines mark the start and the end of
satellite slew. The inset shows the first {\rm group of peaks} at 50~ms time
resolution.  Note that rate units are unchanged but peaks are higher because of
the finer time resolution. The lower panel gives the hardness curve computed
as the ratio of the count rate in the 25$-$150~keV and in the 15$-$25~keV band.
}
\label{batlc2}
\end{figure}

The BAT event data were re-analyzed using the standard BAT analysis software 
(Swift2.4) as described in the Swift BAT Ground Analysis Software Manual 
\citep{krimm}. This incorporates post-launch updates to the 
BAT response and to the effective area and includes the systematic error 
vector to be applied to the spectrum.
The ground analysis of BAT data gave a refined position of the burst
at RA$_{\rm J2000} = 21 ^{\rm h} 23 ^{\rm m} 31\fs8$, 
Dec$_{\rm J2000} = -53\degr 02\arcmin 04\farcs4$, with 
an error radius of $1\arcmin$ (90\% containment, including systematics; \citealp{gcn5256}).
Mask-weighted BAT light curves were created in the standard 4 energy bands
(15$-$25, 25$-$50, 50$-$100, 100$-$150 keV), and in the total 15$-$150 keV 
band (Figs.~\ref{batlc} and \ref{batlc2}) at 1~s time resolution.
These light curves show an unusual multi-peaked burst structure 
that begins with an initial bright {\rm group of peaks} with a 5~s FWHM 
duration, followed by a set of fainter and somewhat softer peaks that increase
in intensity. At about $T_0$+60~s, the light curve shows a hint of 
a last faint peak and then decays smoothly. 
Sub-pulses on the timescale of tens of ms are present, especially 
during the {\rm initial 5$-$8 s} (see the inset in Fig.~\ref{batlc2}).
The duration of the burst can be estimated as $T_{90} = 102 \pm 5$~s, 
but there is observable signal over the interval $T_{100} = 178 \pm 5$~s
with more than 70~s of very faint extended emission after the end of 
the $T_{90}$ time interval.
Since XRT started its observation at full timing and spectral resolution
in windowed timing (WT) mode at T$_0$+97~s, 
there is a $\sim$ 80~s overlap between the BAT and XRT observations.

\begin{figure*}[ht]
\centerline{\includegraphics[width=13cm,angle=-90]{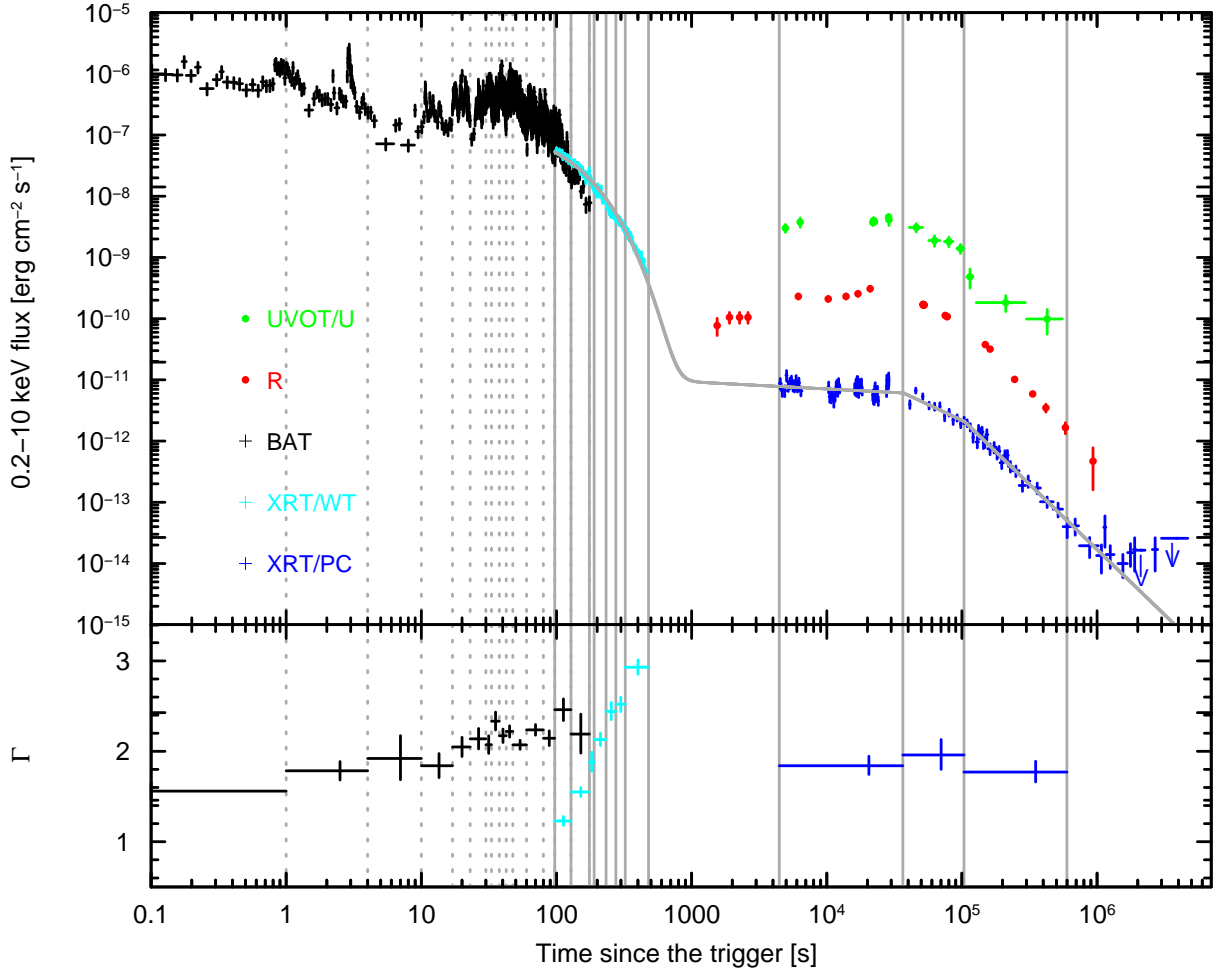}}
\caption{Upper panel: the XRT light curve converted to flux in the 0.2$-$10
keV  energy range is shown together with the BAT light curve extrapolated to
the same  energy range.  The joint best fit of the BAT and XRT spectra in the
time interval 97$-$175 s  has been used to calculate conversion factors for the
BAT and XRT WT mode data. The best power-law fit of the average PC spectrum was
used to convert XRT data in PC mode. The XRT light curve best fit model (solid
line) is over-plotted on the data. The dashed vertical lines show the time
intervals used for extraction of the BAT spectra (Table~\ref{batspecfit}),
while the solid vertical lines show the time intervals used for extraction of
XRT spectra  (corresponding to break times in the light curve during PC
observation, i.e. $t > 1000$~s). The $R$-band light curve (with the host galaxy
contribution subtracted) is shown in arbitrary flux units for comparison. This
light curve is mainly composed of VLT data \citep{dellavalle06} and
complemented with other data available in the literature
\citep{fynbo06,galyam06,gcn5257}.
The $U$-band light curve by UVOT is also plotted  in arbitrary flux units
(with the possible host galaxy contribution subtracted). 
Lower panel: plot of the photon index of BAT and XRT spectra as a function of time
(see Tables \ref{batspecfit}, \ref{specfitwt}, and \ref{specfitpc}).
Note that during the BAT and XRT overlap time interval, the BAT and XRT  photon indices
correspond to the high energy and low energy branches of a Band model \citep{band}, 
respectively (see results of BAT and XRT joint fit in Table \ref{jointspecfit1}).
}
\label{batxrtlc}
\end{figure*}

BAT spectra were extracted over the $T_{100}$ time interval (from $T_0-1.55$~s
to $T_0+176.5$~s, first row in Table~\ref{batspecfit}), 
over the time interval corresponding to the Konus-Wind observation of the first 
{\rm group of peaks} of the burst (from $T_0-2.83$~s to $T_0+5.62$~s), 
over the time interval of simultaneous BAT/XRT observation (from $T_0$+97~s to $T_0$+176.5~s), 
and for the central part of the burst emission (from $T_0$+5.62~s to $T_0$+97~s). 
The last three spectra are indicated in Table \ref{batspecfit} (first column) 
as BAT-A, BAT-C and BAT-B, respectively.
Moreover, BAT spectra were extracted over the 16 time intervals listed in 
Table \ref{batspecfit} to allow for time resolved spectral analysis.
These time intervals are shown by vertical dotted lines in 
Fig.~\ref{batxrtlc}.

Response matrices were generated with the task {\tt batdrmgen} 
using the latest spectral redistribution matrices. 
For each spectrum, relevant keywords for response matrix generation 
were updated with the {\tt batupdatephakw} task.
For our spectral fitting (XSPEC v11.3.2) we considered the 14$-$150\,keV 
energy range and applied the latest energy-dependent systematic error 
vector provided by the {\tt CALDB} distribution\footnote{http://heasarc.gsfc.nasa.gov/docs/swift/analysis/bat\_digest.html}.

\subsection{XRT}
\label{section:xrtanalysis}


The XRT data were first processed by the Swift Data Center at NASA/GSFC 
into Level 1 products (event lists). Then they were further processed 
with the XRTDAS (v1.8.0) software package, written by the 
ASI Science Data Center (ASDC) and distributed within
FTOOLS to produce the final, cleaned event lists. 
In particular, we ran the task {\tt xrtpipeline} (v0.10.3) applying calibration 
and standard filtering and screening criteria. 
For our analysis we selected XRT grades 0$-$12 and 0$-$2 for photon counting (PC) and 
WT data, respectively (according to Swift nomenclature; \citealt{bur05a,hill04}). 
The X-ray counterpart was detected at the position 
RA$_{\rm J2000} =21^{\rm h} 23^{\rm m} 32\fs00$,  
Dec$_{\rm J2000} =-53^{\circ} 01^{\prime} 39\farcs4$, 
with an estimated uncertainty of $3\farcs7$.
This position was determined using the {\tt xrtcentroid} task (v0.2.7) 
on the PC data in sequence 000, that are not affected by pile-up, 
and it takes into account the correction for the misalignment
between the telescope and the satellite optical axis.
It is $51\farcs7$ from the refined BAT position.
 
\subsubsection{First orbit data}

During sequence 000 the count rate of the burst was high enough 
to cause pile-up in the WT mode data, that covered 
the entire first orbit XRT observation from T$_0$+97 to T$_0$+480~s.
Therefore, to account for this effect, the WT data were 
extracted in a rectangular 40$\times$20-pixel region 
with a 9$\times$20-pixel region excluded from its centre.
The size of the exclusion region was determined following the 
procedure illustrated in \citet{romano06}: the analysis  of
the fraction of events at grade 0 for different sizes of the
central hole in the extraction region saturates to a constant
value for sizes greater than 10 pixels.
To account for the background, WT events were also extracted within a 
rectangular box (40$\times$20 pixels) far from background sources. 

\begin{figure}[hb]
\includegraphics[width=6.2cm,angle=-90]{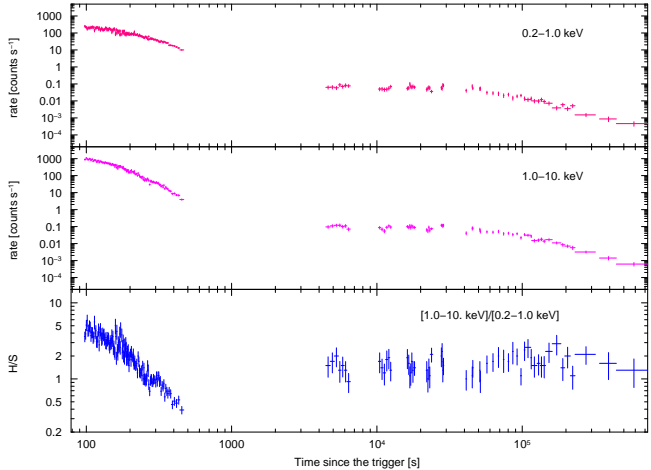}
\caption{XRT light curves in the 0.2$-$1.0~keV (upper panel) and 1.0$-$10~keV
(middle panel) energy bands and hardness ratio curve (lower panel).}
\label{hardness}
\end{figure}

Background subtracted WT light curves were extracted in the
0.2$-$1.0~keV, 1.0$-$10~keV and 0.2$-$10~keV energy ranges
(see Figs. \ref{batxrtlc} and \ref{hardness}).
They were corrected for both the fraction of point spread function 
(PSF) lost due to the central hole in the extraction region 
and exposure variations within the extraction region. 
A vignetting correction was also properly applied.

Time-resolved spectral analysis of WT data was performed on 7 time intervals
(see Table~\ref{specfitwt} and Fig.~\ref{batxrtlc})
selected according to the source brightness
and requiring at least 2000 net counts each. 
The time resolved spectra were extracted from a rectangular 
40$\times$20-pixel region with a central rectangular region excluded. 
Different sizes of the excluded region were used according to
the analysis of the fraction of events at grade 0 (see details
in Table~\ref{specfitwt}). 
Note that spectra WT-1 and WT-2 are simultaneous to spectra
BAT-15 and BAT-16, respectively.

An average WT spectrum 
and a spectrum simultaneous to the tail of the BAT light curve 
(i.e. simultaneous to the BAT-C spectrum) 
were also extracted. 
The former is labeled as WT in the first column of Table~\ref{specfitwt}, 
and the latter is labeled as WT-0. 
Both were extracted using 
the 10 pixels wide central hole 
to correctly account for the highest degree of pile-up 
at the beginning of the observation.

\subsubsection{Data taken after 1000 s}

From the second orbit on, XRT observed in PC mode.
For the PC data, which were never affected by pile-up throughout all the
XRT observations, we extracted the source events in a circular 
region of 30 pixel radius up to segment 003 (i.e. $t\sim 556$~ks).
PC background data were also extracted in a source-free  
circular region (radius 40 pixels), and the background subtracted 
light curves (in the 0.2$-$1.0~keV, 1.0$-$10~keV and 0.2$-$10~keV 
energy ranges) were corrected for the fraction of PSF lost, 
for time-dependent exposure variations within the extraction region 
and for the vignetting effect.

For sequences from 004 to 037 the light curve points
were calculated using the task {\tt sosta} of the {\tt ximage} 
package, which calculates vignetting- and PSF-corrected count rates 
within a specified box, and the background in a user-specified region. 
The background was estimated in the same region as the one used 
for the initial part of the light curve. 
Starting from sequence 006, data segments were merged until 
source detection with signal to noise ratio larger than 3 was attained
(see Table \ref{log}).
The last detection shown in Fig.~\ref{batxrtlc} at 2.7~Ms, having 
a signal to noise ratio of 1.8, may be a statistical fluctuation.

Three PC spectra were extracted according to different evolutionary
stages in the light curve (see \S~\ref{section:batxrtlc}).
They are listed in Table~\ref{specfitpc}.
Ancillary response files were generated with the task {\tt xrtmkarf} 
within FTOOLS, and account for different extraction regions and 
PSF corrections. 
The basic files describing the XRT effective area used were the 
{\tt swxpc0to12\_20010101v008.arf} (for PC) and 
{\tt swxwt0to2\_20010101v008.arf} (for WT)
from the latest distribution of the XRT Calibration Database 
(CALDB 2.4) maintained by HEASARC.
We also used the latest spectral redistribution matrices in the 
Calibration Database, namely the file {\tt swxpc0to12\_20010101v008.rmf} 
for PC spectra and the file {\tt swxwt0to2\_20010101v008.rmf} for WT spectra.


\subsection{UVOT}
\label{section:uvotanalysis}

The first UVOT observation was a 97 s White finding chart taken 
in IMAGE\&EVENT mode starting 104 s after the BAT trigger \citep{gcn5255}.  
In this first observation the UVOT located the optical afterglow 
of GRB~060614 at RA$_{\rm J2000}=21^{\rm h} 23^{\rm m} 32\fs08$, 
Dec$_{\rm J2000}=-53^{\circ}01^{\prime} 36\farcs2$
with a 90\% confidence interval of $0\farcs56$.  
This is 3$\farcs$3 from the centre of the XRT error circle.

To extract UVOT light curves we performed aperture photometry on the UVOT
exposures using a circular aperture with a radius of $2\arcsec$
centred on the UVOT position of the afterglow.  A sky annulus of width
$7\farcs5$ and inner radius $27\farcs5$ was used.  This annulus
includes a large number of sky pixels and is large enough to exclude
the faint outer regions of the point-spread function of the afterglow.
We performed aperture corrections to convert the $2\arcsec$ aperture
photometry to the standard photometric apertures used to define the
UVOT photometric zero points ($6\arcsec$ for {\sl UBV\/} and
$12\arcsec$ for the UVW1, UVM2, UVW2, and White filters).
Approximately between 5 and 7 isolated stars (depending on the filter) 
were used to compute aperture corrections for each exposure.  
The RMS scatter in the aperture
corrections for each source in a single exposure is typically 0.02 mag.
The instrumental magnitudes were transformed to Vega magnitudes
using the photometric zero points in the {\sl Swift\/}/UVOT
calibration database (CALDB).  Colour terms have not been applied, but
preliminary calibrations suggest that they are negligible for sources
with typical afterglow colours.  The adopted photometric zero points
are ZP$_V = 17.88 \pm 0.09$, ZP$_B = 19.16 \pm 0.12$, ZP$_U = 18.38
\pm 0.23$, ZP$_\mathrm{UVW1} = 17.69 \pm 0.02$, ZP$_\mathrm{UVM2} =
17.29 \pm 0.23$, ZP$_\mathrm{UVW2} = 17.77 \pm 0.02$, and
ZP$_\mathrm{White} = 19.78 \pm 0.02$.

Figure~\ref{uvotlc} shows the Swift-UVOT UVW2, UVM2, UVW2, $U$, $B$, 
$V$ and White light curves, together with the $R$-band light curve  of the
afterglow mostly composed by VLT data \citep{dellavalle06} and complemented
with points taken from the literature \citep{fynbo06,galyam06,gcn5257}. The
flux densities are monochromatic fluxes for the central wavelength  of each
filter. They have been corrected for the \citet{shlegel98} 
Galactic extinction along the line of sight to GRB~060614 ($A_V=0.07$~mag). 

The afterglow of GRB~060614 becomes brighter at optical and ultraviolet (UV) 
wavelengths until approximately eight hours after the BAT trigger 
and then fades.  There is weak evidence for {\rm small-timescale 
fluctuations around the average light curve} between approximately 
3 and 28 hours.  
This may be due to energy injection during the first day after the burst; 
however, due to the photometric uncertainty in the data, and the inherent 
difficulties with co-adding images of a variable source, it is not clear 
if these fluctuations are real. {\rm We note that the $R$-band
light curve (as observed by ground based telescopes) shows small-timescale
variability as well.} The $U$ and UVM2 light curves show 
weak evidence for change in the decay rate at approximately 28 hours after
the BAT trigger.  

The UVOT $U$-band detection 
above the power law fit to the light curve at $\sim 500$~ks
was previously  interpreted as a possible SN \citep{gcn5281,gcn5286}.
In view of the extremely low limits on a SN contribution obtained by
ground-based and HST observations \citep{dellavalle06,fynbo06,galyam06}, 
we now interpret this excess as a statistical fluctuation.
The shape of the excess is hard to determine due to the faintness 
of the afterglow and the subsequent exposure time (spanning days) 
needed to detect it, and is dependent on how the power-law is
{\rm fitted} to the earlier data and how the late time data are co-added.  
We present here the most conservative version, consisting of a 
single detection about five days after the burst which lies above 
the power-law fit, with the later data co-added together 
to determine the host galaxy brightness. 
%


\section{Data analysis}
\label{section:analysis}

\subsection{The BAT and XRT light curves}
\label{section:batxrtlc}

The XRT light curve in 0.2$-$10 keV flux units is shown
in Fig.~\ref{batxrtlc} together with the BAT light curve
extrapolated to the XRT energy band. 
Both light curves have been converted to flux using
the best fit model parameters of the joint fit of the
BAT and XRT spectra extracted from the BAT/XRT overlap
time interval 97$-$176.5 s 
(spectra BAT-C and WT-0, respectively; see \S~\ref{section:specres}). 
Note that the best fit model to the BAT and XRT spectra
in the overlap time interval, as presented in \S~\ref{section:specres},
is a Band model \citep{band} with peak energy $E_{\rm p}\sim 8$~keV, 
and the extrapolation of the BAT light curve to the XRT energy range 
according to this model naturally leads to a very good match between 
the tail of the BAT and the start of the XRT light curve,
but it underestimates slightly the 0.2$-$10 keV emission of 
the burst before the start of the XRT observation, when the
peak energy of the spectrum was certainly higher 
(as high as $\sim 300$ keV during the first {\rm episode} in BAT).
However, a time-dependent conversion of the BAT light curve 
to the XRT energy range cannot be performed because of our
ignorance about the exact $E_{\rm p}$ value after the {\rm initial
group of BAT peaks} and before the start of the XRT observation.

The XRT/WT light curve can be modeled as an exponential
decay, while the XRT/PC light curve is well {\rm fitted} by a
doubly broken power law model. Results of the light curve fit
are presented in Table~\ref{xrttimefit}. Note that points 
at $t > 1.8$~Ms slightly deviate from the best fit model 
and suggest a late re-brightening or a flattening of the X-ray 
afterglow. 
However, the last detection at 2.7~Ms is only a 1.8-$\sigma$
detection, and the following upper limit tells us that at the
end the source faded below the XRT sensitivity limit.

A fit with an additional constant (to account for the 
late flattening seen in the light curve) gives a slightly
steeper slope $\alpha_{\rm C}=2.2\pm0.1$, and a constant flux 
of $\sim(8\pm4)\times10^{-15}$~erg~cm$^{-2}$~s$^{-1}$.
This would correspond to an X-ray luminosity of about 
$3\times10^{41}$ erg~s$^{-1}$. 
This value is by far larger than what can be provided by the ongoing star
formation activity: using the star formation rate $1.3 \times
10^{-2}~M_\odot~\mathrm{yr}^{-1}$ \citep{dellavalle06} and the conversion
factor by \citet{grimm03}, this could contribute $\la 10^{36}$ erg s$^{-1}$. 
The X-ray luminosity could however be due to a small AGN.
A further study of the optical spectrum of the galaxy mighty clarify 
this issue. We note however that \citet{galyam06}  show that the 
light profile of the galaxy is well {\rm fitted} by an exponential disk model, 
without the need of a nuclear component.
Therefore, the apparent flattening of the late X-ray light curve
may also be a statistical fluctuation.

The 0.2$-$1.0~keV and 1.0$-$10~keV XRT light curves are shown
in the first two panels of Fig.~\ref{hardness}. They have been 
rebinned so as to have at least 100 counts per bin in each energy band 
during  the first orbit (WT data) and at least 20 counts per bin 
in each energy band afterwards (PC data). 
A hardness ratio curve was computed and it is shown in the third panel.
A strong hard to soft evolution is visible in the first 400 s of data.
This is likely due to the passage of the peak energy across the
XRT band (see \S \ref{section:specres}).

Note that portions of smooth bumps are visible during the nearly flat
light curve segment from 4 to $\sim30$~ks and the hardness curve
tracks them, although the hardness variations are small.
Then, the light curve shows a hardening trend up to the time of the second
break $T_{\rm b,2}\sim104$~ks, and an irregular behaviour with
oscillating hardness later on. However, the hardness curve at $t>4$~ks
is well {\rm fitted} by a constant with best fit value 1.42$\pm0.05$ 
($\chi^2=44.52$, with 56 d.o.f.).


\begin{figure}[ht]
\centerline{\includegraphics[width=19cm,angle=-90]{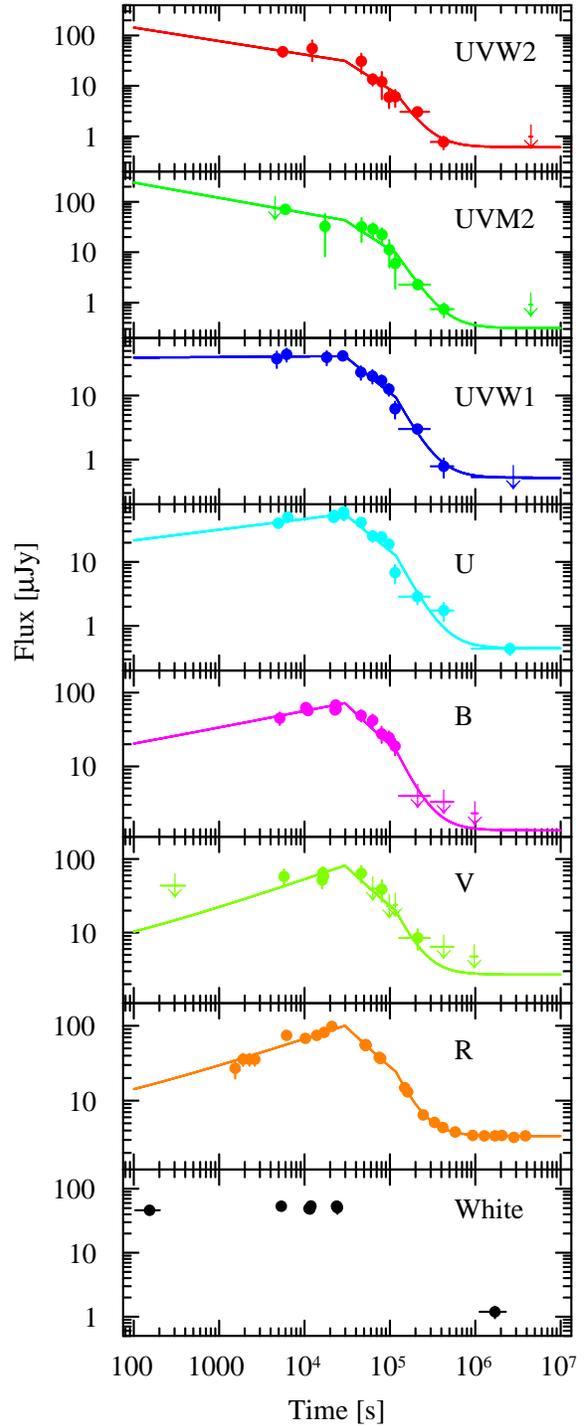}}
\caption{UVOT light curves in different filters and $R$-band light curve
obtained from VLT data \citep{dellavalle06}
and other data in the literature \citep{fynbo06,galyam06,gcn5257}.
All light curves are corrected for the Galactic extinction along the 
line of sight to GRB~060614.
The best fit model discussed in the text is represented by solid lines.
}
\label{uvotlc}
\end{figure}

\subsection{The UVOT light curves}
\label{section:uvotlc}

The Swift-UVOT UVW2, UVM2, UVW2, $U$, $B$, $V$ and White 
light curves shown in Fig.~\ref{uvotlc} are too sparse for a detailed 
filter-dependent fit. On the other hand, the $R$ band light curve of the afterglow 
obtained using data from the literature \citep{dellavalle06,fynbo06,galyam06,gcn5257},
shown in the seventh 
panel from the top in Fig.~\ref{uvotlc} 
can be well {\rm fitted} by a constant (accounting for the host galaxy emission) 
plus a doubly broken power law.
We performed a joint fit to the UVOT and $R$-band light curves using the same 
model, with filter-dependent initial slope ($\alpha_{\rm A}$),
host galaxy contribution ($h$), and normalization. 
The values of $h$ for the filters $B$, $V$ and $R$ were fixed to measured values: 
$B=23.73 \pm 0.13$, $V=22.75\pm 0.05$, and $R=22.46 \pm 0.01$ \citep{dellavalle06}.
The White filter UVOT light curve, shown in the bottom 
panel of Fig.~\ref{uvotlc}, 
does not sample uniformly the observation time interval and has not been included 
in the fit. 
The best fit parameters are listed in Table \ref{uvottimefit}.

We find that the behaviour of all our optical and ultraviolet 
light curves is consistent with a first break at $T_{\rm b,1}$=29.7$\pm$4.4 ks,
an after-break decay with slope  $\alpha_{\rm B}$=1.11$\pm$0.05,
a second break at $T_{\rm b,2}$=104$\pm$22.0 ks, and a final decay
with slope $\alpha_{\rm C}$=2.44$\pm$0.08.
The decay slope before the first break ($\alpha_{\rm A}$) 
has a well defined trend with wavelength (decreasing from
the ultraviolet to the optical), although errors are quite large.
We note that, on the contrary, the light curve in the White filter is constant
up to 30 ks.

The best fit values of the host galaxy contributions $h$ in the ultraviolet
filters (UVW2, UVM2, UVW1, $U$) are roughly consistent with those of a dwarf
star-forming galaxy.


\setcounter{table}{4}
\begin{table*}[ht]         
 \begin{center}         
 \caption{Results of XRT time-resolved analysis: PC data.}     
 \label{specfitpc}
 \normalsize
 \begin{tabular}{lllllll} 
 \hline 
 \hline 
 \noalign{\smallskip} 

        Spectrum  &          Start time        &         End time   &          $N_{\rm H}$$^{\mathrm{a}}$           & 
        $\Gamma$  &          d.o.f. & $\chi^2_{\rm r}$ \\
                  &          (s)               &         (s)        &          ($10^{20}$~cm$^{-2}$) & 
                  &                            \\

 \noalign{\smallskip} 
 \hline 
 \noalign{\smallskip}

  PC-1 & 4448.6   &  36575.4    & $<$1.9              & 1.84$_{-0.095}^{+0.105}$   &~66 & 1.08  \\
  \noalign{\smallskip}                                                                                                                                      
  PC-2 & 36575.4  &  103701.8   & $<$3.9              & 1.96$_{-0.16 }^{+0.17 }$   &~25 & 0.96  \\
  \noalign{\smallskip}                                                                                                                                      
  PC-3 & 103701.8 &  598638.2   & $<$2.9              & 1.77$_{-0.11 }^{+0.12 }$   &~51 & 0.83  \\

  \noalign{\smallskip} 
  \hline 
  \noalign{\smallskip} 

  PC   & 4448.6   &  598638.2 & 1.5$_{-1.2 }^{+1.2 }$ & 1.84$_{-0.08 }^{+0.08 }$ &  146 & 0.99\\ 

  \noalign{\smallskip}
  \hline
  \end{tabular}
  \end{center}
  \begin{list}{}{} 
  \item[$^{\mathrm{a}}$]Extragalactic absorption column. The model included
                         both a {\tt wabs} component accounting for Galactic absorption (i.e. with $N_{\rm H}$ 
                         fixed to $3\times10^{20}$~cm$^{-2}$; \citealp{Dickey1990}) and a {\tt zwabs} component
                         with redshift fixed to $z=0.125$ and free $N_{\rm H}$ parameter to account for
                         extragalactic absorption.
  \end{list} 
  \end{table*} 


\subsection{BAT and XRT spectral analysis}
\label{section:specres}

All BAT spectra are best {\rm fitted} by single power law models.
Results are shown in Table \ref{batspecfit}.
The BAT average spectrum of the burst has a photon index $\Gamma=2.13 \pm 0.04$, 
corresponding to a 15$-$150 keV fluence of 2.2$\times 10^{-5}$ erg cm$^{-2}$. 
A fit with a Band model \citep{band} with low energy index $\alpha_{\rm Band}$ 
fixed to $-1$ does not represent an improvement, but allows us to set an upper limit 
to the average peak energy of $E_{\rm p} \simlt 24$~keV at the 90\% confidence 
level.
The BAT spectrum of the first {\rm episode of peaks} 
(spectrum BAT-A in Table~\ref{batspecfit})
has a photon index of 1.63$_{-0.07 }^{+0.07 }$,
consistent with the Konus-Wind low energy index 1.57$_{-0.14 }^{+0.12 }$ \citep{gcn5264}.
The average spectrum of the following five broad peaks (spectrum BAT-B) is softer
($\Gamma=2.2\pm0.04$).  The BAT spectrum of the tail of the prompt emission
(spectrum BAT-C) has a similar photon index, $\Gamma=2.37\pm0.13$. 
The 15$-$150 keV fluence in the {\rm initial group of peaks} 
is 3.4$\times 10^{-6}$ erg cm$^{-2}$,
while the rest of the burst provides a fluence of  1.9$\times 10^{-5}$ erg cm$^{-2}$.
The softening of the BAT emission is well represented in the lower panel of
Fig.~\ref{batxrtlc}, where the best fit photon indices of the 16 BAT spectra 
are plotted as a function of time.

The XRT WT spectra have also been modeled with an absorbed single power law
(see Table \ref{specfitwt}, top panel). 
All the fits were done in the 0.3$-$10 keV energy range excluding channels 
between 0.45 and 0.55 keV where an instrumental artefact commonly
appears in XRT spectra\footnote{This artefact is probably caused
  either by a time-dependent energy offset, or by a problem in the
  detector response matrix, and is under active investigation by the
  XRT instrument team; see details in
  http://www.swift.ac.uk/xrt\_bias.pdf\,.}.
Two absorption components were included in the model: a Galactic absorption
component, fixed to the expected value $N^{\rm G}_{\rm H} =
3\times10^{20}$~cm$^{-2}$ according to \citet{Dickey1990}, and an absorption
component intrinsic to the host which was left free to vary.
The simple power law fits implied intrinsic absorption at the level of 
$\sim 5 \times10^{20}$~cm$^{-2}$ in all of the XRT/WT spectra 
but showed an unphysical trend in the $N_{\rm H}$ consisting of a rise 
followed by a decay.
Moreover, the reduced $\chi^2$ of the fits are marginally
acceptable and careful inspection of the fit residuals put in evidence
systematic trends below 2 keV that suggest the presence of curvature 
in the spectra.

To account for the curvature of the WT spectra we fit them also with
{\it i)} an absorbed cut-off power law model, {\it ii)} an absorbed
Band model \citep{band}, and
{\it iii)} an absorbed power law model plus a blackbody component.
The choice of the first two models is justified by the possibility that
the early XRT light curve is the tail of the prompt emission, which usually
has a Band spectrum (the Band spectrum can be approximated by a
cut-off power law over a limited energy band). 
Model {\it iii)} was suggested by recent results obtained by \citet{campana06} 
on GRB~060218 and \citet{grupe06} on GRB~060729.
Best fit parameters of all XRT/WT spectra with these three models
are shown in Table~\ref{specfitwt}.
All the three models gave us a significant improvement in the fit
and are statistically equivalent. They give total reduced $\chi^2$
(calculated over the WT-1 to WT-7 spectra) of 1.02 (990 d.o.f),
1.01 (988 d.o.f.), and 1.01 (983 d.o.f.), respectively.
They also constrain intrinsic absorption to the level 
of a few$\mbox{} \times 10^{20}$~cm$^{-2}$ or less.
In the model {\it (iii)} fits, 
the temperature of the blackbody component $kT_{\rm bb}$ smoothly decreases from 0.8 
to 0.2~keV, while the photon index $\Gamma$ grows from 1.3 to 2.8.
The radius of the emitting region is of the order of $10^{11}$~cm and seems to peak
at $\sim T_{0} + 250$~s, as does the fraction of total flux contributed by the blackbody 
component.
The fits with models {\it i)} and {\it ii)}, on the other hand, give us the
alternative picture of a smooth peak energy decrease from $\sim$ 8 keV 
to a value below the XRT energy range at about $\sim T_{0} + 250$~s.

\begin{figure}[ht]
\centerline{\includegraphics[width=6cm,angle=-90]{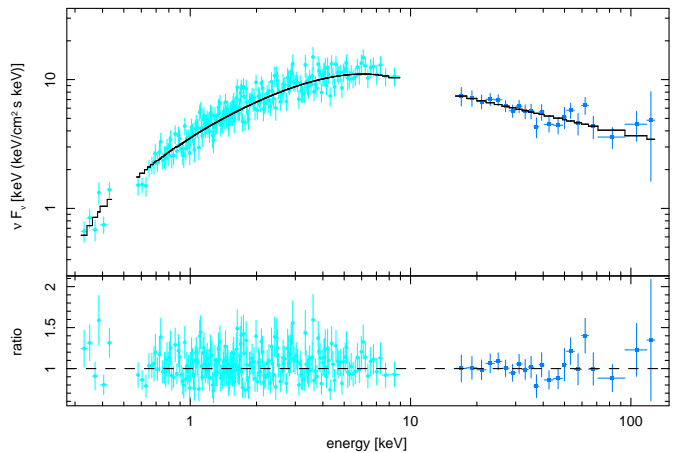}}
\caption{Joint fit of the BAT and XRT spectra in the BAT/XRT overlap time 
interval. The best fit model is {\rm an absorbed} Band model with peak 
energy $\sim8$~keV. See Table~\ref{jointspecfit1}.}
\label{batxrtspec}
\end{figure}


A clue to the true nature of the spectrum is found by a simultaneous
fit to the BAT and XRT spectra during the overlap time when both
instruments were detecting the burst 
(spectral pairs BAT-C and WT-0; BAT-15 and WT-1; BAT-16 and WT-2).
When fitted with power laws, the BAT and XRT spectra show substantially different 
photon indices during these intervals, implying the existence of a break in the 
broad band spectrum.
A joint fit of these pairs of spectra was therefore performed with 
an absorbed Band model. Results are shown in Table~\ref{jointspecfit1}
(top panel)
and in Figure~\ref{batxrtspec}.
A peak energy $E_{\rm p}$ of 8~keV is detected during the BAT/XRT
overlap (Fig. \ref{batxrtspec}).
A description of the joint BAT/XRT spectra with an absorbed power 
law model plus a blackbody is statistically less successful and 
requires an absorption column ten times higher than all previous fits
(bottom panel of Table~\ref{jointspecfit1}).
This argues against the presence of a blackbody component in the 
early XRT spectra, though we cannot rule it out after 176~s 
post-burst.


The PC spectra (starting after 4~ks from the trigger) are well {\rm fitted} 
by a single power law model and did not show evidence of intrinsic absorption. 
Results are in Table \ref{specfitpc}.
The low absorption values are consistent with those
found for the WT data fit to intrinsically curved spectral models,
adding additional evidence that the power law fits to the WT data are inappropriate.
No significant spectral variation is seen throughout the XRT/PC observation.
The three spectra extracted, corresponding to the three decay phases observed
after 4 ks, show photon indices consistent with the average value of 
1.85$\pm$0.12.  


\setcounter{table}{5}
 \begin{table}[hb]         
 \begin{center}         
 \caption{GRB~060614 X-ray light curve best fit parameters.}     
 \label{xrttimefit}
 \normalsize
 \begin{tabular}{llc} 
 \hline 
 \hline 
 \noalign{\smallskip} 
 Parameter  &   &  value \\
   \noalign{\smallskip} 
\hline
  \noalign{\smallskip} 

$\tau$                    &  (s)          & 75.7  $\pm$ 0.4   \\
$\alpha_{\rm A}$          &               & 0.11  $\pm$ 0.05  \\ 
$T_{\rm b,1}$             &  (10$^3$ s)   & 36.6  $\pm$ 2.4   \\ 
$\alpha_{\rm B}$          &               & 1.03  $\pm$ 0.02  \\ 
$T_{\rm b,2}$             &  (10$^3$ s)   & 103.7 $\pm$ 22.0  \\ 
$\alpha_{\rm C}$          &               & 2.13  $\pm$ 0.07  \\ 
$\chi^{2}$ (d.o.f.)       &   (WT)        & 319 (183) \\ 
$\chi^{2}$ (d.o.f.)       &   (PC)        & 119 (~97) \\ 

  \noalign{\smallskip} 
  \hline
  \end{tabular}
  \end{center}
  \begin{list}{}{} 
  \item Note. The model used to fit the XRT light curve consisted of an exponential decay law
        ($\propto \exp{-t/\tau}$) plus a doubly-broken power-law.
        The parameters $\alpha_{\rm A}$, $\alpha_{\rm B}$, and $\alpha_{\rm C}$ are the decay slopes 
        for the distinct phases of the afterglow. 
        $T_{\rm b,1}$ and $T_{\rm b,2}$ are the epochs at which the decay slope changes,
        measured from the GRB onset. Contributions to the $\chi^{2}$ of the WT and PC datasets
        have been indicated.
  \end{list} 
  \end{table} 

\setcounter{table}{6}
 \begin{table}[ht]         
 \begin{center}         
 \caption{GRB~060614 optical light curve best fit parameters.}     
 \label{uvottimefit}
 \normalsize
 \begin{tabular}{lllc} 
 \hline 
 \hline 
 \noalign{\smallskip} 
 Parameter  &  Filter  &  & Value \\
   \noalign{\smallskip} 
\hline
  \noalign{\smallskip} 

$h$                     & UVW2  &  (mag)        & 23.25 $\pm$  0.54  \\ 
$h$                     & UVM2  &  (mag)        & 24.1  $\pm$  1.0 \\ 
$h$                     & UVW1  &  (mag)        & 23.5  $\pm$  0.7 \\ 
$h$                     & $U$   &  (mag)        & 23.8  $\pm$  0.3 \\ 
$h$                     & $B$   &  (mag)        & 23.73   \\ 
$h$                     & $V$   &  (mag)        & 22.75   \\ 
$h$                     & $R$   &  (mag)        & 22.46   \\ 
$\alpha_{\rm A}$        & UVW2  &               &  0.27 $\pm$   0.26 \\ 
$\alpha_{\rm A}$        & UVM2  &               &  0.30 $\pm$   0.26 \\ 
$\alpha_{\rm A}$        & UVW1  &               & -0.01 $\pm$   0.20 \\ 
$\alpha_{\rm A}$        & $U$   &               & -0.17 $\pm$   0.14 \\ 
$\alpha_{\rm A}$        & $B$   &               & -0.23 $\pm$   0.15 \\ 
$\alpha_{\rm A}$        & $V$   &               & -0.41 $\pm$   0.55 \\ 
$\alpha_{\rm A}$        & $R$   &               & -0.38 $\pm$   0.23 \\ 
$T_{\rm b,1}$           & all   &  (10$^3$ s)   & 29.7   $\pm$  4.4  \\ 
$\alpha_{\rm B}$        & all   &               & 1.11   $\pm$  0.05 \\ 
$T_{\rm b,2}$           & all   &  (10$^3$ s)   & 117.2  $\pm$  4.4  \\ 
$\alpha_{\rm C}$        & all   &               & 2.44   $\pm$  0.08 \\ 
$\chi^{2}$ (d.o.f.)     &       &               & 68.815 (74) \\ 

  \noalign{\smallskip} 
  \hline
  \end{tabular}
  \end{center}
  \begin{list}{}{} 
  \item Note.
        The UVOT light curves and the $R$-band light curve of the GRB~060614 afterglow
        were fitted simultaneously with a constant plus a doubly-broken power law model.
        The parameter $h$ represents the constant value of the host galaxy contribution
        to the total observed flux. The $h$ values are not corrected for any extinction.
        The parameters $\alpha_{\rm A}$, $\alpha_{\rm B}$, and $\alpha_{\rm C}$ are the decay slopes 
        for the distinct phases of the afterglow. 
        $T_{\rm b,1}$ and $T_{\rm b,2}$ are the epochs at which the decay slope changes,
        measured from the GRB onset. During the fit only $h$, $\alpha_{\rm A}$ and the 
        normalizations were allowed to be filter dependent.
  \end{list} 
  \end{table} 

\setcounter{table}{7}
 \begin{table}[ht]         
 \begin{center}         
 \caption{Results of BAT and XRT joint spectral analysis.}     
 \label{jointspecfit1}
 \normalsize
 \begin{tabular}{llll} 
 \hline 
 \hline 
 \noalign{\smallskip} 

BAT Spectrum        &                     BAT-C              &        BAT-15                         &         BAT-16   \\  
XRT Spectrum        &                     WT-0               &        WT-1                           &         WT-2     \\  
Time$^{\mathrm{a}}$ &                     97--176.5          &                97--128                &         128--175             \\
Central hole$^{\mathrm{b}}$ &             10                 &         10                            &         7                    \\
                                                                                                                                                           
  \noalign{\smallskip}                                                                                                                                     
\hline                                                                                                                                                     
  \noalign{\smallskip}  
\multicolumn{4}{c}{Band model} \\
  \noalign{\smallskip} 
\hline
  \noalign{\smallskip} 
                                                                                                                                                           
$N_{\rm H}$$^{\mathrm{c}}$  &     $<$0.7                     &       $<$0.6                          & $<$0.9                      \\ 
  \noalign{\smallskip}                                                                                                                                     
$\alpha_{\rm Band}$$^{\mathrm{d}}$ &   -0.89$_{-0.08}^{+0.05}$ &      -0.86$_{-0.11}^{+0.05}$        & -0.84$_{-0.13}^{+0.07}$     \\
  \noalign{\smallskip}                                                                                                                                     
$\beta_{\rm Band}$$^{\mathrm{d}}$  &   -2.38$_{-0.09}^{+0.07}$ &      -2.46$_{-0.06}^{+0.12}$        & -2.10$_{-0.17}^{+0.13}$     \\
  \noalign{\smallskip}                                                                                                                                     
$E_{\rm p}$$^{\mathrm{e}}$ &      6.1$_{-0.7}^{+1.3}$        &      8.3$_{-1.2}^{+2.5}$              & 4.5$_{-1.1}^{+1.4}$         \\
  \noalign{\smallskip}                                                                                                                                     
(d.o.f.)                   &      359                        &            240                        & 298                         \\  
  \noalign{\smallskip}                                                                                                                                     
$\chi^2_r$          &             0.99                       &       1.00                            & 0.97                        \\  

  \noalign{\smallskip} 
\hline
  \noalign{\smallskip} 
\multicolumn{4}{c}{Power-law model plus blackbody}\\
  \noalign{\smallskip} 
\hline
  \noalign{\smallskip} 

$N_{\rm H}$$^{\mathrm{c}}$    & 13.3$_{-2.5 }^{+2.9}$    & 10.9$_{-3.1 }^{+3.8}$  &  3.0$_{-0.8 }^{+6.0}$    \\
  \noalign{\smallskip} 
$\Gamma$                      & 2.1$_{-0.1 }^{+0.1}$     & 2.3$_{-0.1}^{+0.1}$    &  1.5$_{-0.1}^{+0.1}$     \\ 
  \noalign{\smallskip} 
$kT_{\rm bb}$$^{\mathrm{f}}$  & 1.29$_{-0.08 }^{+0.09}$  & 1.31$_{-0.09}^{+0.09}$ &  0.7$_{-0.1}^{+0.2}$     \\
  \noalign{\smallskip} 
$R_{\rm bb}$$^{\mathrm{g}}$   & 1.4$_{-0.2 }^{+ 0.2}$    & 1.9$_{-0.3 }^{+ 0.3}$  &   2.2$_{-0.7 }^{+ 0.7}$  \\   
  \noalign{\smallskip} 
d.o.f.                        & 359                      &  240                   &   298                    \\   
  \noalign{\smallskip}  
$\chi^2_r$                    & 1.27                     &  1.23                  &   1.10                   \\   

  \noalign{\smallskip} 
  \hline
  \end{tabular}
  \end{center}
  \begin{list}{}{} 
  \item[$^{\mathrm{a}}$] Start and stop times of the time interval in seconds.
  \item[$^{\mathrm{b}}$] Length in pixels of the central box (20 pixel thick) that we excluded 
                         from the $40\times20$ pixel extraction region to account for the pile-up effect in XRT/WT data.
  \item[$^{\mathrm{c}}$] Extragalactic absorption column in units $10^{20}$~cm$^{-2}$. The model included
                         both a {\tt wabs} component accounting for Galactic absorption (i.e. with $N_{\rm H}$ 
                         fixed to $3\times10^{20}$~cm$^{-2}$; \citealp{Dickey1990}) and a {\tt zwabs} component
                         with redshift fixed to $z=0.125$ and free $N_{\rm H}$ parameter to account for
                         extragalactic absorption.
  \item[$^{\mathrm{d}}$] $\alpha_{\rm Band}$ and $\beta_{\rm Band}$ are the low and high energy index of the Band model, respectively.
  \item[$^{\mathrm{e}}$] Peak energy of the Band model in units of keV. 
  \item[$^{\mathrm{f}}$] Black body temperature in units of keV.
  \item[$^{\mathrm{g}}$] Black body radius in units of $10^{11}$~cm. 
  \item Note. The first column in the table corresponds to the BAT and XRT/WT spectra integrated over the whole                                             
              BAT/XRT overlap time interval.
  \end{list} 
  \end{table} 

\subsection{X-ray/optical spectral energy distributions}
\label{section:seds}

\begin{figure*}[ht]
\centerline{\includegraphics[width=8cm,angle=0]{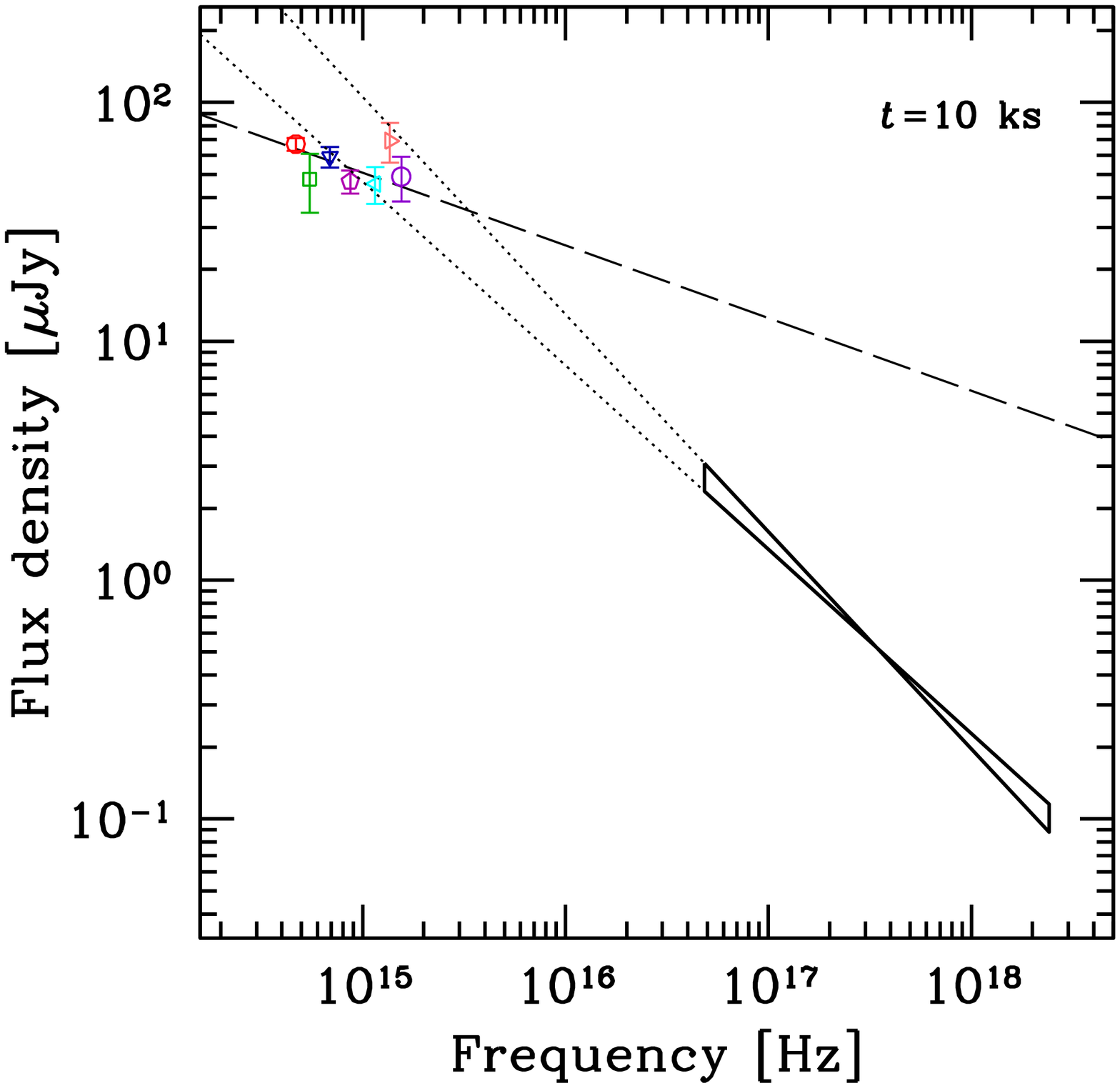},\includegraphics[width=8cm,angle=0]{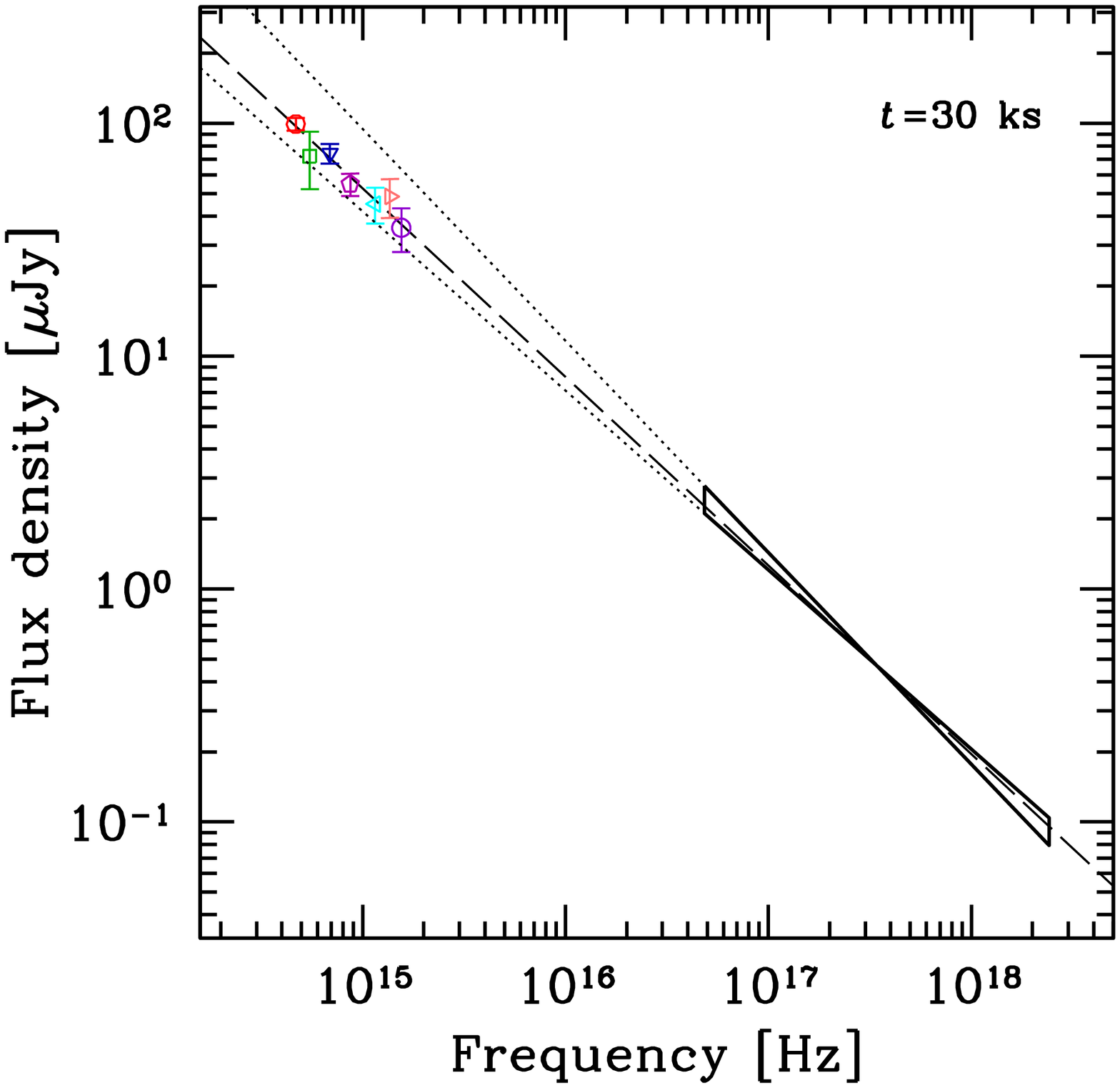}}
\centerline{\includegraphics[width=8cm,angle=0]{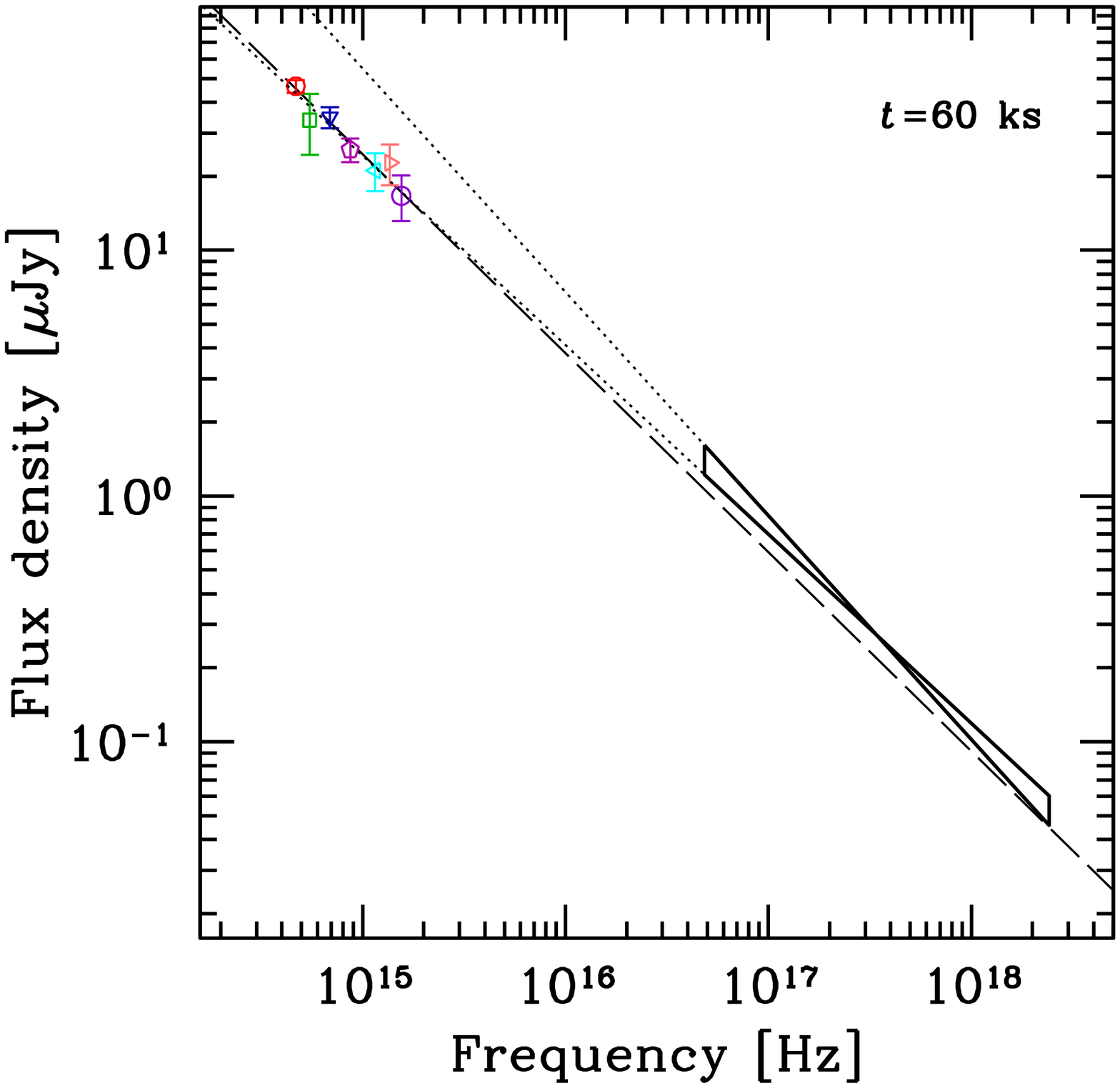},\includegraphics[width=8cm,angle=0]{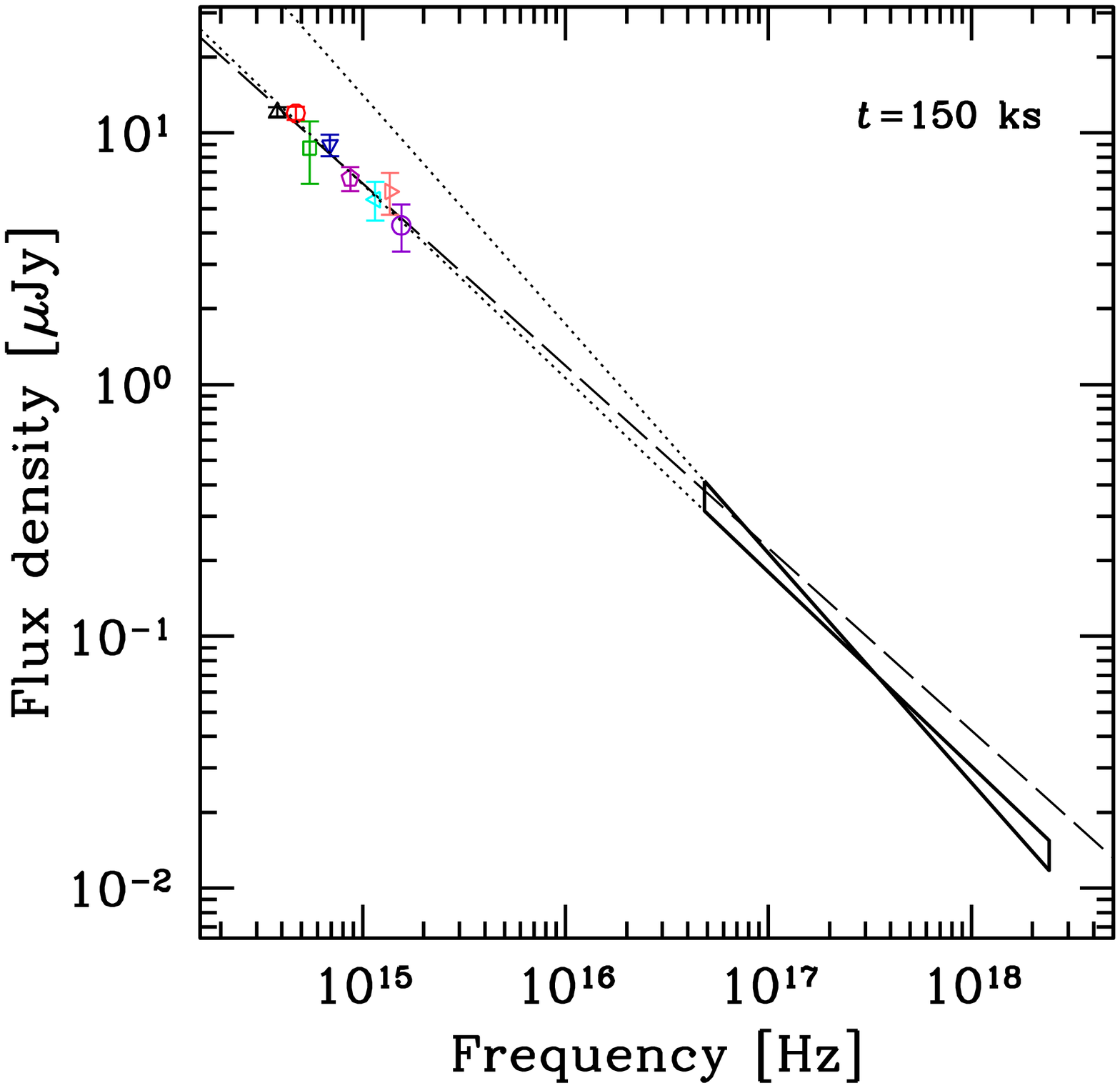}}
\caption{Optical, ultraviolet and X-ray spectral energy distribution  at 10,
30, 60 and 150 ks after the trigger. Optical and UV data, shown as points, 
have been corrected for the Galactic extinction along the line of sight
($A_V=0.07$~mag) and for a host extinction of $A_V=0.05$~mag with an SMC-like
extinction curve. The SED at 150 ks also contains the $I$ magnitude of the
afterglow obtained by VLT observations \citep{dellavalle06}. The solid lines
define the cone corresponding to the 90\% uncertainty on the spectral slope
in the X-ray band. The extrapolation of the cone to the optical band is shown
by the dotted lines. The dashed lines represent the best fit to the optical/UV
data.}
\label{sedplot}
\end{figure*}

The spectral energy distribution (SED) of the afterglow 
from optical to X-rays have been computed at several different times
using the best fit models of the UVOT and VLT 
light curves presented in \S~\ref{section:uvotlc}, the best fit model 
of the XRT light curve (\S~\ref{section:batxrtlc}) and the average 
XRT spectrum of the PC observation (\S~\ref{section:specres} 
and Table \ref{specfitpc}).
We selected the following times as representative of the afterglow 
evolutionary stages: 10, 30, 60 and 150 ks, i.e. a time before
the first break/peak, the time of the first break/peak,
a time between the first and the second break and a time
after the second break. 
The SED at 150 ks also contains the $I$ magnitude of the afterglow 
extrapolated from VLT observations \citep{dellavalle06}.
Infrared, optical and UV data have been corrected for the Galactic extinction
along the line of sight ($A_V=0.07$) and for a host extinction of
$A_V=0.05\pm0.02$ with an SMC-like extinction curve. 
The value for the host galaxy extinction was computed in order 
to provide the best match for the last three SEDs.
Results are shown in Fig. \ref{sedplot}. 
The SED at 10~ks implies a spectral break between the X-ray and
optical bands.  We note that since the early optical light curves are different at different frequencies
the spectrum is changing, and $\beta_{\rm opt}$ gets redder with time for 
$t < T_{\rm b,1}$.
After $T_{\rm b,1}$ all the SEDs are consistent with the
optical and X-ray data belonging to the same power-law segment. 
Since the PC data do not show spectral evolution, the slope and 
opening of the cone representing the spectral distribution in the X-ray band 
and its uncertainty are constant and given by $\beta_{\rm X}=0.84\pm0.07$.
The best fit of the optical and UV data has a slope of 
$\beta_{\rm opt}=0.30 \pm 0.14$ at 10 ks, and $\beta_{\rm opt}=0.81\pm0.08$ 
at later times. 
So $\Delta \beta = \beta_{\rm X}-\beta_{\rm opt} = 0.54 \pm 0.16$ at  10 ks, 
and  $\Delta \beta = 0.03\pm 0.11$ at later times.
%


\section{Results and Discussion}
\label{section:results}

\subsection{Prompt emission and early X-ray light curve}
\label{section:prompt}

The GRB~060614 prompt emission consists of an initial short hard 
{\rm emission structure with many bright peaks} 
($\sim$5~s FWHM duration, $E_{\rm p}\sim 300$~keV) 
followed by a longer ($\sim 170$~s), softer and {\rm highly variable} 
bump {\rm which ends in a} tail that smoothly matches the early and 
partly simultaneous X-ray light curve observed by XRT. 

\begin{figure}[ht]
\centerline{\includegraphics[width=8cm,angle=0]{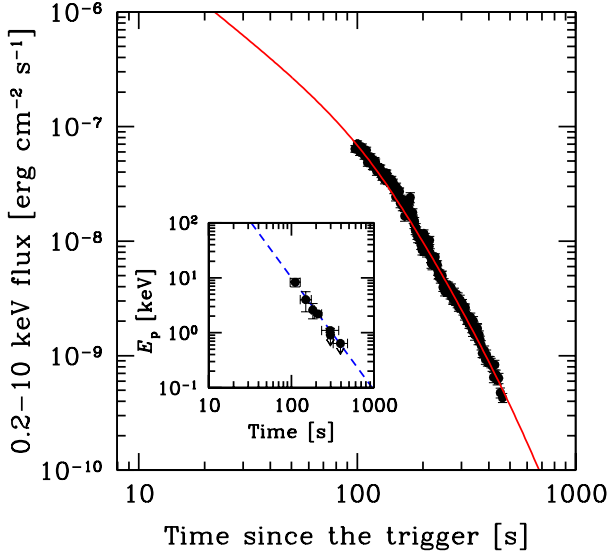}}
\caption{Modeling of the early XRT light curve (black points) as the 
emission of a Band spectrum with evolving peak energy (red solid line).
The $E_{\rm p}$ evolution law used is the best fit of the measured 
$E_{\rm p}$ values obtained through time resolved spectral analysis and
presented in Table \ref{specfitwt}. The $E_{\rm p}$ decay law with time
is shown in the inset (blue dashed line).}
\label{bandlcmodel}
\end{figure}

The XRT light curve shows strong spectral evolution during the
very steep decay up to the end of the first orbit ($\sim 500$~s), 
then enters a nearly flat decay phase followed by steeper decay phases 
that will be discussed later in \S \ref{section:closure}. 
We now focus only on data before 500~s.
The X-ray light curve over this time interval is not a power-law. 
In Table \ref{xrttimefit}, we report fit results assuming 
an exponential shape, adopted for sake of simplicity only. 
The spectral fits suggest a physical interpretation of this steepening decline. 
The combined BAT/XRT spectrum is in fact well described by a Band
model (or a cutoff power law) with decreasing peak energy $E_{\rm p}$. 
The passage of $E_{\rm p}$ through the XRT band would naturally produce a
steepening in the light curve, since the flux density at $E > E_{\rm p}$ 
decreases faster than at $E < E_{\rm p}$. The width of the XRT bandpass 
(0.2$-$10 keV) will make the transition smooth when considering 
the band-integrated flux. 
This behaviour is represented in Fig.~\ref{bandlcmodel}, where we show 
the expected 0.2$-$10 keV light curve for a Band spectrum with decreasing 
peak energy and normalization. 
We fixed the low and high-energy photon indices to $\alpha_{\rm Band}=-1.1$ 
and $\beta_{\rm Band}=-2.6$ (see third panel of Table \ref{specfitwt}). 
The time behaviour of $E_{\rm p}$ was obtained by interpolating the 
observed values (Fig.~\ref{bandlcmodel}, inset), which show a regular 
decay $E_{\rm p} \propto t^{-\alpha}$ with $\alpha = 2.04 \pm 0.11$. 
The agreement with the observed data is satisfactory, so that 
the light curve shape provides further support for the spectral 
evolution pattern described in \S~\ref{section:specres}. 
The observation of $E_{\rm p}$ passing through the XRT band during early XRT
observations is indeed a natural possibility {\rm \citep{butler06}}, being
consistent with  the well-known hard-to-soft evolution observed in the prompt
emission  spikes {\rm \citep{ford95}}.
Then, the measured $E_{\rm p}$ variation strongly supports the interpretation 
of the early XRT light curve of GRB~060614 as the low energy counterpart 
of the fading and softening tail seen by BAT. 
According to the analysis of \citet{zhangbb07},
the GRB~060614 tail behaviour cannot be explained as the simple superposition
of high latitude radiation \citep{kumar00,dermer04} and a possible prolonged 
and steady central engine emission \citep{zhang04}, but may be explained 
as the result of the cooling frequency decrease associated to adiabatic cooling 
of shock heated shells after an internal collision.

In their combined analysis of BAT and XRT data of Swift GRBs,
\citet{obrien06} and \citet{willingale06} concluded that all X-ray light 
curves can be well described by an exponential that relaxes into a power law, 
often with flares superimposed. 
There have also been a few cases with clearly detected exponential phases 
in the early XRT light curve (e.g. \citealp{vaughan06}).
By analogy with GRB\,060614, we suggest that also for these cases the light curve 
was shaped by the passage of the peak energy inside the XRT band. Hardness ratio
plots can be used to test this hypothesis as done in the case of XRF~050416A 
by \citet{mangano07} to show the consistency of Swift observations with
a decaying $E_{\rm p}$ scenario.

The alternative time-dependent fit of early XRT data of GRB~060614
with the power law plus blackbody model 
(fourth panel in Table \ref{specfitwt}) may be unphysical,
particularly given the fact that the broad band spectrum from 
97 to 176~s is clearly {\rm fitted} by a Band function, which may well 
continue up to 500~s. 
The blackbody fit may be just a computational 
way to fit the curvature in a Band spectrum peaking within 
the narrow XRT energy range, and the observed decrease of 
the blackbody radius (related the normalization of the blackbody
component) may be associated to the exit of the peak energy from the XRT
energy band.

However, it is worth noticing that in GRB~060614 the long bump 
of the prompt {\rm emission} peaks at about 40$-$50~s since the trigger.
According to \citet{kobazhang06}, in the thick shell scenario 
the onset of the afterglow (and the corresponding peak
in the light curve) would occur at about the fireball
deceleration time
$T_{\rm dec}=(3 E / 32 \pi \rho c^5 \Gamma_0^8)^{1/3}$,
where $E$ is the fireball total energy, $\rho$ is the mass density 
of the surrounding medium,
$\Gamma_0$ is the initial Lorentz factor of the expanding fireball
and $c$ is the speed of light. 
If we assume a radiation efficiency of shocks $\eta\sim0.2$
we can estimate $E=E_{\rm iso}/\eta$ with
$E_{\rm iso}\sim 2.5 \times 10^{51}$~erg  
(see \S~\ref{section:energetics}).
Under the further assumptions $\rho\sim 10 m_{\rm p}$~cm$^{-3}$
(with $m_{\rm p}$ being the proton mass) 
we obtain $T_{\rm dec}\sim 45$~s for $\Gamma_0\sim 100$.

On the other hand, the putative blackbody component observed after 100~s 
from the trigger has a temperature $kT_{\rm bb}\sim 0.9$~keV and 
radius $R_{\rm bb}\sim 2\times 10^{11}$~cm. This is similar to 
what expected from the slowly expanding ($\sim 10000$~km~s$^{-1}$) 
SN shock front after break-out from the stellar surface
if the SN radiation energy budget is of the order of $10^{51}$~erg. 
Then, the radiation we observe may be the superposition of  standard internal
shock activity up to 100 s from the trigger, afterglow onset and blackbody
radiation from expanding SN shock.   In this scenario, the absence of SN
detection in the optical could be explained by a very under-luminous SN
explosion due to fall back over the nuclear region of heavy elements (mainly
Ni) synthetized in the core collapse {\rm \citep{nomoto04} or  by a ``dark
hypernova'' from a metal-poor massive star  \citep{nomoto07,tominaga07}. Given
the deep optical upper limits \citep{galyam06,dellavalle06,fynbo06}, this would
be by far the most under-luminous SN associated to a GRB ever detected, and one
of the faintest SNe ever discovered.} 


\subsection{Energetics}
\label{section:energetics}

An accurate estimate of the isotropic equivalent energy radiated 
by GRBs requires the knowledge of their intrinsic average spectrum over 
the 1 keV$-$10~MeV energy band in the source rest frame.
The narrow  BAT sensitivity bandpass (15$-$150 keV)
makes it hard to detect GRB peak energies in most cases.
The average spectrum of GRB~060614 is well {\rm fitted} by a single 
power-law with photon index $\simgt 2$. This indicates that we are observing
the high energy branch of the Band law generally used to describe GRBs 
spectra, and a 24 keV upper limit to the average peak energy can be set 
in the observer frame (see \S~\ref{section:specres}). Moreover, we know 
that the burst presented strong spectral evolution and the peak energy 
decreased from $\sim 300$~keV measured during the {\rm initial group of peaks} 
to $\sim 8$~keV during the BAT/XRT overlap. Then we expect that 
the average peak energy of the burst should be larger than 8 keV.
With a peak energy in the 8$-$24~keV range, GRB~060614 is definitely
an X-ray rich burst and may even be an X-ray flash \citep{lamb05}. 
A lower limit to the isotropic equivalent energy $E_{\rm iso}$ can be
obtained under the extreme assumption that the GRB emission below 
$24\times(1+z)$ keV in the source rest frame is zero. An upper
limit can be obtained under the assumption that the peak energy is
below 1~keV (i.e. extrapolating the $\Gamma \sim 2.13$ power law 
to low energies). In this way we obtain 
$1.8\times 10^{51}~\mathrm{erg} < E_{\rm iso} < 3.2 \times 10^{51}$~erg, 
and can reasonably assume 
$E_{\rm iso}=(2.5\pm0.7)\times 10^{51}$~erg.
According to this estimate and the limits on the average peak
energy discussed above, GRB~060614 is consistent at about 2 sigma
with the $E_{\rm p}^{\rm rest} - E_{\rm iso}$ correlation found
by \citet{amati} for long GRBs with known
redshift and recently updated including a larger sample of events 
(\citealt{newamati}; see also \citealp{amati06}).

The rest-frame peak isotropic luminosity in the 30$-$10000 keV range
{\rm (calculated assuming the broad band spectrum given by Konus-Wind,
\citealp{gcn5264})} 
is $L_{\rm p,iso}= 5.3_{-1.8}^{+1.0} \times 10^{49}$ erg s$^{-1}$.
This value is consistent with the extrapolation to low luminosities
of the $E_{\rm p}^{\rm rest}$-$L_{\rm p,iso}$ relation found by
\citet{yonetoku04}.

The isotropic equivalent energy radiated during the {\rm initial} hard 
{\rm episode of peaks} 
$E_{\rm iso,1p}$ can also be estimated. The best fit photon index 
of the BAT spectrum of the {\rm initial group of peaks}, 
$\Gamma \sim 1.6$, is in good agreement with the best fit photon index 
of Konus-Wind data over the same energy range and indicates 
that the peak energy was above the BAT energy range.
Based on the 302~keV peak energy detected by Konus-Wind, we obtain $E_{\rm
iso,1p} \sim 3.5 \times 10^{50}$~erg. {\rm Then, only about one seventh of the
isotropic energy  was released during the initial group of peaks. As noted also
by \citep{gehrels06}, } the first hard {\rm episode} of GRB~060614 is not 
consistent with the Amati correlation. {\rm This is not surprising, since the
Amati correlation was computed using global, and not time-resolved,
properties.}


\subsection{Breaks and closure relations}
\label{section:closure}

The XRT light curve for $t > 4$~ks shows  
an initial flat decay with slope $\alpha_{\rm A}=0.11\pm0.05$  
and a constant spectral energy index $\beta_{\rm A}=0.84\pm0.10$, 
a steepening to a slope $\alpha_{\rm B}=1.03\pm0.02$ 
at 36.6$\pm$2.4~ks without significant spectral evolution 
($\beta_{\rm B}=0.96\pm0.16$) and a final steepening to a slope 
$\alpha_{\rm C}=2.13\pm0.07$ at 104$\pm$22~ks without significant 
changes in the spectrum ($\beta_{\rm C}=0.77\pm0.12$). 
This behaviour resembles the typical behaviour of Swift detected
X-ray afterglows \citep{nousek06,obrien06}.
The slow decay rate of phase A can be interpreted as the effect of forward
shock refreshing mechanisms, likely ending at the time of the first break 
\citep{zhang06,nousek06,panaitescu06}.
The slope of the phase B decay is intermediate between the value expected in
the cases $\nu_{\rm X} < \nu_{\rm c}$ and $\nu_{\rm X} > \nu_{\rm c}$
($\nu_{\rm c}$  being the synchrotron cooling frequency), namely  $\alpha =
3\beta/2 = 1.26\pm0.10$ and $\alpha = 3\beta/2-1/2 = 0.76 \pm 0.10$.  Since
phase B lasted for a relatively short time (only half a decade),  the value of
$\alpha_{\rm B}$ may be affected by systematic errors  due to the choice of the
fitting function (we chose for simplicity a piecewise-connected power law,
while the real shape of the transition will be smooth; \citealp{granot01}). The
hard spectral shape suggests that $\nu_{\rm c}$ is above the X-ray band.

The striking feature of the multi-band light curve, however, is the presence of
an achromatic break at $T_{\rm b,2}$, observed simultaneously in the XRT light
curve as well as in the UVOT and $R$-band ones. An achromatic transition is
expected when the edge of the jet enters the visible portion of the emitting
surface. Note that \citet{dellavalle06} found a break in the optical afterglow
light curve at $119\pm3.4$~ks with pre-break slope $1.08\pm0.03$ and post-break
slope $2.48\pm0.07$. Then, our simultaneous fit of the UVOT light curves with
the $R$-band light curve from \citet{dellavalle06} points toward the
consistency of all the optical/UV data with a break at $117.2 \pm 4.4$ ks. 
This confirms the jet-break nature of the second break  of our X-ray light
curve.  The X-ray and optical slopes after the break are roughly consistent
with each other, and consistent with the decay predicted after a jet break.
{\rm To our knowledge, this is one of the best examples of a jet break for a
Swift burst.}
A jet-break at $\sim$104~ks implies a jet opening angle of $\vartheta_{\rm
jet}$ = $0.161~[t_{\rm b}/(1+z)]^{3/8}(n~\eta_{\gamma}/E_{52})^{1/8}$~rad
\citep{sari99} where, $t_{\rm b}$ is the break time in days, $E_{52}$ is the
isotropic energy release in units of $10^{52}$~erg, $n$ is the particle density
of the circumburst matter in cm$^{-3}$ and $\eta_{\gamma}$ is the conversion 
efficiency of internal energy to gamma-rays. For an isotropic
energy $E_{\rm iso} \sim 2.5 \times 10^{51}$~erg we obtain  $\vartheta_{\rm jet} =
10.5^{\circ}\, (n/3)^{1/8}(\eta_{\gamma} / 0.2)^{1/8}$. 
This corresponds to a beaming-corrected energy  
$E_{\gamma} = E_{\rm iso}(1-\cos \vartheta_{\rm jet})\sim 4.2 \times 10^{49}$~erg 
and makes GRB~060614 roughly consistent with the  $E_{\rm p}^{\rm rest} -
E_{\rm \gamma}$ relation originally found by \citet{ghir} and re-investigated
by \citet{liang05} and \citet{GHIRLANDA_2007}.
Note that the beaming-corrected energy of GRB~060614 
is lower than typical $E_{\gamma}$ of long GRBs ($\sim 10^{51}$~erg)
and comparable to the beaming-corrected energy of the short burst 
GRB~051221A \citep{burrows06}.

The interpretation of the first break of the X-ray light curve as a
hydrodynamical break due to the end of forward shock refreshing mechanisms
would also require an achromatic break. Indeed, the optical and ultraviolet
light curves of the afterglow show another common break at about 30 ks, nearly
simultaneous with the first X-ray break (that occurs at about 37 ks). However,
the initial slope $\alpha_{\rm A}$ is dependent upon wavelength at optical
frequencies. This implies that the spectrum was changing at these wavelengths
around $t \sim 10$~ks. In particular, since the UV light curves are initially
decaying, while at lower frequencies a clear rise is observed, the spectrum is
evolving from blue to red.
A way to explain this behaviour is the passage of a break frequency through the
optical/UV band. This kind of transition should also be accompanied by a light
curve slope change as well, but is not achromatic. The observed behaviour of
all our afterglow light curves could be accounted for if the break frequency
passed through the observed band slightly earlier than the hydrodynamical
break. In this scenario the observed initial break/peak in the UVOT and VLT
light curves would be due to the superposition of the two effects (causing the
change in the spectrum) while the initial break in the X-ray light curve would
be purely hydrodynamical (with no spectral evolution across it, as observed).
Which kind of break frequency can produce the observed behaviour? The simplest
possibility is that we were observing the injection frequency $\nu_{\rm i}$,
that is the frequency at which the lowest-energy electrons are radiating. 
For $\nu < \nu_{\rm i}$, where the spectrum is very blue ($F_\nu \propto
\nu^{+1/3}$), rising light curves are expected, while a standard decay occurs
for $\nu > \nu_{\rm i}$. As the injection frequency moves to longer wavelengths, 
the curves above $\nu_{\rm i}$ would be decaying, while the curves below 
$\nu_{\rm i}$ would be still rising. The very blue spectrum is confirmed
by our spectral analysis of the early-time SED 
(which has $\beta_{\rm opt} = 0.30 \pm 0.14$). 
If this interpretation is correct, this could be the first time for which the
injection frequency was directly observed in the optical band.

The other possibility is that the change in the spectral shape around 
$t = 10$~ks is due to the passage of the cooling frequency. 
For this model to work, however, the shocked electron power law slope 
should be $p = 2 \beta_{\rm X} \sim 1.7$. 
Using the formulae of \citet{daicheng01} and \citet{zhang04} to
predict the light curve slope in case of forward shock  propagation in the slow
cooling regime in a uniform interstellar medium when $p < 2$, we obtain
$\alpha_{B}$ between 0.9 and 1, roughly consistent with the observation. 
The main problem of this interpretation is that the required $p$ value (1.7) is
substantially lower than the value consistent with the jet break interpretation
of the second break (2.2). A temporal evolution of the shocked electron power
law slope within the first 30 ks could reconcile the models. 
A $p$ change, i.e. a shock microphysics change, during an  afterglow evolution is
not an implausible event and could be related to changes in the dynamics of the
shock like those associated with the end of shock refreshing mechanisms.
This would however produce a change in the X-ray spectrum as well, 
which is not observed.

{\rm
\subsection{Comparison with other bursts}

\label{section:comparison}

\begin{figure*}[ht]
\centerline{\includegraphics[width=13cm,angle=-90]{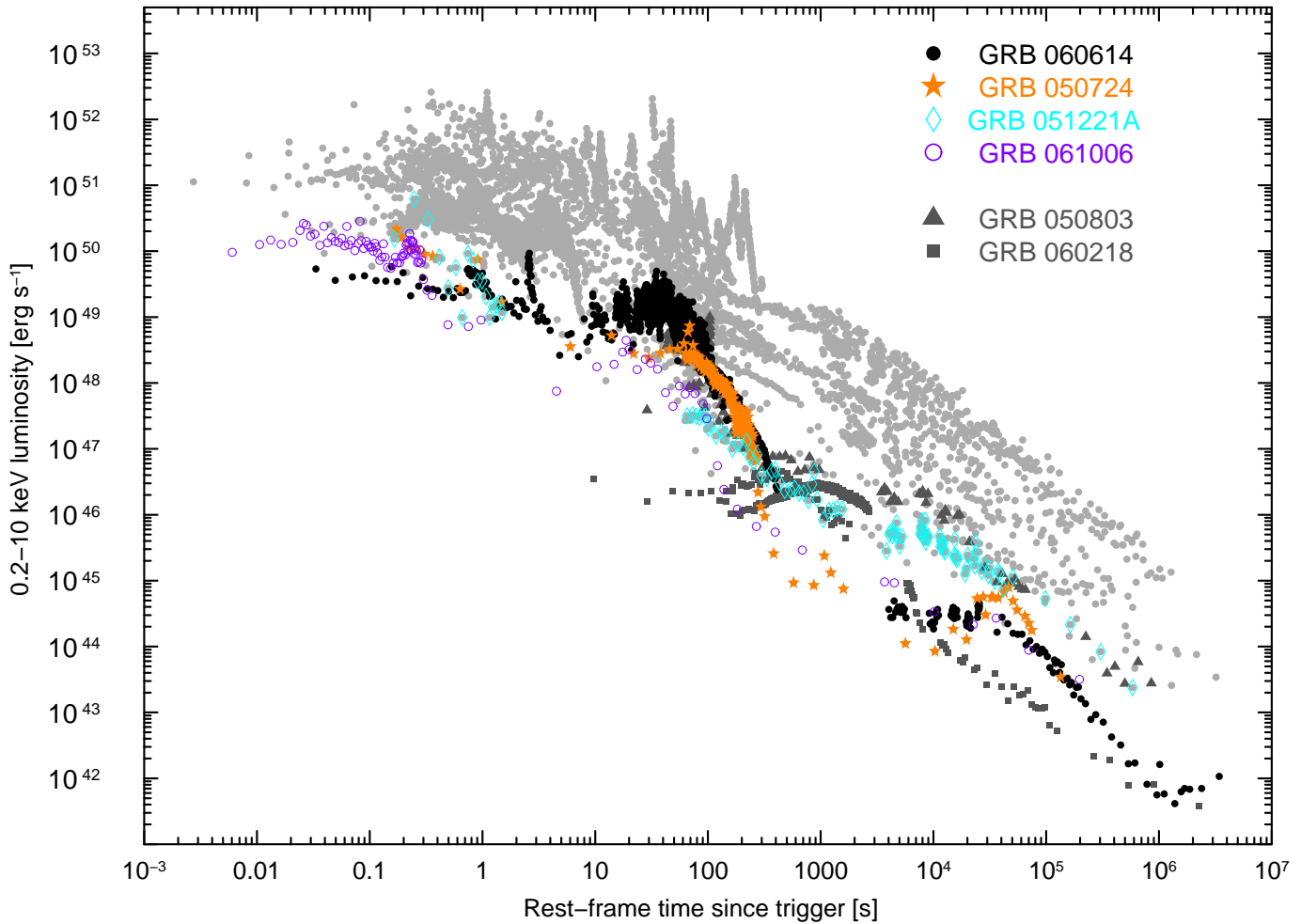}}
\caption{\rm Rest-frame luminosity light curves of several GRBs detected by
Swift: GRB~060614 (black points), the 3 short-duration GRBs with secure
redshift (GRB~050724, GRB~051221A, and GRB~061006) and 37 long bursts with 
known redshift (light and dark grey curves). The only two long GRBs with redshift
z$<0.5$ are evidenced (dark grey curves). Note that the sample contains other 
5 long bursts with $0.5 < {\rm z} < 1$. Each light curve consists
of {\it i)} the BAT light curve of the prompt emission extracted in the
15$-$150 keV energy band, extrapolated to flux in the 0.2$-$10 keV energy band
using the best fit parameters of the BAT spectrum and then converted into
0.2$-$10 keV luminosity using the appropriate $k$-correction; {\it ii)} the XRT
light curves, converted to 0.2$-$10 keV luminosity with the appropriate
$k$-correction required for the best fit model of the XRT spectra. See
\citet{mangano07b} for further details.}
\label{shortlong}
\end{figure*}

Despite the contrasting properties of GRB~060614, which make it difficult to
classify unambiguously, its afterglow does not show anomalous features.
Phenomenologically, its X-ray light curve shows all the ``canonical''
components identified in long GRB afterglows
\citep{nousek06,obrien06,willingale06}. Furthermore, its properties are in
remarkable agreement with the predictions of jetted fireball models
\citep{sapina98,sari99}.
However, several recent studies have outlined that the X-ray afterglows of
short GRBs show the same basic features as well:
power law decay with changing slopes,  X-ray flares (GRB~050724:
\citealt{campana06b,grupe06b}; GRB~051210: \citealt{laparola06}), shallow decay
phases (GRB~060313: \citealt{roming06}; GRB~061006: \citealt{grb061006report}),
breaks (GRB~061201: \citealt{grb061201report}; GRB~061006:
\citealt{grb061006report}) and possibly jet breaks (GRB~051221A:
\citealt{burrows06}). This makes it difficult to infer conclusions about the
nature of GRB~060614 based on its light curve properties.
An interesting feature observed in a fraction of short-duration GRBs is the
existence of a soft hump following the initial short, hard spike \citep{norrisbonnel06}.
As noted by several authors \citep{gehrels06,zhang07}, this feature resembles
what observed in GRB~060614, which may be an extreme case of such behaviour.

In Fig. \ref{shortlong} we plotted the (rest-frame) luminosity light curves of
several GRBs detected by Swift, including GRB~060614, the 3 short GRBs with
secure spectroscopic redshift\footnote{We require the existence of an optical
counterpart in order to have a certain identification of the host galaxy.}
(GRB~050724, GRB~051221A, GRB~061006), and 37 long GRBs with measured redshift
(mostly from \citealt{mangano07b}).

The light curve of GRB~060614 is strikingly similar to those of GRB~050724 and
GRB~061006, in terms of both shape and luminosity. Note that these two bursts
showed an extended soft emission episode
\citep{barth05b,campana06b,grb061006report}. We note, however, that the shape
of the GRB~060614 light curve is quite similar to that of several long-duration
GRBs. In terms of luminosity, taken at a face value, it looks that GRB~060614 
is at the faint end of the distribution, being $\sim 50$ times fainter than 
the average. This comparison, however, is a bit misleading given the big
difference in the (average) redshift of most long-duration bursts and
GRB~060614. Among the plotted objects, only GRB~060218 and GRB~050803 have 
$z < 0.5$ ($z = 0.0331$ and $z = 0.422$, respectively;
\citealt{REF_Z_060218,REF_Z_05083}), and, not surprisingly, they have both
faint light curves, comparable in luminosity to GRB~060614 and to the short
GRBs. Of course, given the steep GRB luminosity function \citep{REF_FIRMANI},
closeby objects are most likely to be faint on the average, given the small
sampled volume at low redshift. It is also whortwhile to mention that one of
the striking features of GRB~060614, namely the lack of a bright associated
supernova, would pass unnoticed at larger redshift, so that the number of
SN-less event is actually poorly constrained. Only the detection of a
statistically significant number of long-duration events at moderate/low
redshift will make it possible a meaningful comparison.

}


\section{Conclusions}
\label{section:conclusions}

GRB~060614 is a very peculiar, nearby ($z=0.125$) burst. 
Despite the long duration of its prompt emission 
($T_{90}=102\pm5$~s), 
it presents null time lags in the BAT bandpass (similar to short bursts)  and
has no evidence of an associated supernova down to very deep limits
($M_{V}\sim-14$). {\rm GRB~060614 lies close to the region occupied by short
bursts in the time-lags/peak-luminosity plane, but, on the other hand,
satisfies the $E_{\rm iso}-E_{\rm p}^{\rm rest}$ (Amati) correlation, that
holds for long bursts only.} Its BAT light curve shows a prolonged softening
tail that smoothly matches the partly simultaneous early XRT light curve. The
temporal decay of this tail is well described by an exponential function, with
an $e$-folding time of $\sim76$~s, and the spectral evolution is associated
with the drift of the peak energy of the Band function towards lower energy and
crossing the XRT energy band.

The X-ray, UV and optical afterglow light curves and SEDs
of GRB~060614 are well interpreted within the standard afterglow theory.
In particular, the afterglow of GRB~060614 shows a truly achromatic
break from optical to X-ray frequencies at $\sim104$~ks after the trigger, 
fully consistent with being a jet break. {\rm The burst satisfies the
$E_{\gamma}-E_{\rm p}^{\rm rest}$ (Ghirlanda) correlation.} 
The observed softening of the optical/UV afterglow
before 30 ks can be interpreted as an evidence of passage of the 
injection frequency through the optical band.

{\rm The good agreement of the afterglow of GRB~060614 with the jetted fireball
models does not allow us to draw firm conclusions on the nature of the event
and its possible progenitor, but is indeed remarkable since many Swift bursts
hardly reconcile with the very same models.}


\begin{acknowledgements}
This work is supported at INAF by funding from 
ASI on grant number I/R/039/04 and from MIUR 
grant 2005025417,
at Penn State by NASA contract NASS5-00136, 
and at the University of Leicester by the Particle
Physics and Astronomy Research Council. 
We gratefully acknowledge the contribution of dozens 
of members of the XRT team at OAB, PSU, UL, GSFC,
ASDC and our sub-contractors, who helped make 
this instrument possible.
DM acknowledges support from the Instrument Center 
for Danish Astrophysics.
\end{acknowledgements}

%

%

\setcounter{table}{3}
\begin{table*}[p]         
 \begin{center}         
 \caption{Results of XRT time-resolved spectral analysis: WT data.}     
 \label{specfitwt}
 \normalsize
 \begin{tabular}{lllllllll} 
 \hline 
 \hline 
 \noalign{\smallskip} 
Spectrum               &          WT         &          WT-1         &          WT-2         &          WT-3          & 
                                  WT-4       &          WT-5         &          WT-6         &          WT-7          \\
Time$^{\mathrm{a}}$    &          ~97--480   &          ~97--128     &          128--175     &          175--190      & 
                                  190--233   &          233--275     &          275--322.5   &          322.5--480    \\
Central hole$^{\mathrm{b}}$ &     10         &          10           &          7            &          5             & 
                                  3          &          2            &          0            &          0             \\
 \noalign{\smallskip} 
 \hline 
 \noalign{\smallskip} 
\multicolumn{9}{c}{Power-law model} \\
  \noalign{\smallskip} 
\hline
  \noalign{\smallskip} 

$N_{\rm H}$$^{\mathrm{c}}$    & 3.6$_{-0.8}^{+0.8}$     & $2.0_{-1.5}^{+2.0}$         &  $7.0_{-2.0}^{+2.0}$        & $10.0_{-3.0}^{+4.0}$       &  $9.3_{-1.9}^{+2.1}$       &  $7.4_{-1.9}^{+2.2}$       &  $5.0_{ -1.1}^{+1.3}$      &  $5.5_{-1.1}^{+1.2}$     \\        
  \noalign{\smallskip} 
$\Gamma$                      & 1.61$_{-0.03 }^{+ 0.03}$ & $1.23_{-0.05}^{+0.05}$     &  $1.55_{-0.05}^{+0.05}$     & $1.88_{-0.10}^{+0.11}$     &  $2.13_{-0.07}^{+0.07}$    & $2.44_{-0.09}^{+0.10}$     & $2.52_{-0.08}^{+0.08}$     & $2.93_{-0.08}^{+0.08}$   \\ 
  \noalign{\smallskip} 
d.o.f.                        & 386                      & 185                        & 243                         & 69                         &      160                   & 96                         & 119                        & 125                      \\ 
  \noalign{\smallskip} 
$\chi^2_{\rm r}$              & 1.19                     & 1.11                       &  1.23                       & 1.11                       &      1.32                  & 1.48                       & 1.03                       & 1.17                     \\ 

  \noalign{\smallskip} 
\hline
  \noalign{\smallskip} 
\multicolumn{9}{c}{Power-law model with exponential cut off} \\
  \noalign{\smallskip} 
\hline
  \noalign{\smallskip} 

$N_{\rm H}$$^{\mathrm{c}}$    & $<$0.9                  & $<$0.6                   & $<$1.0                    &  $3.6_{-3.3}^{+4.4}$       &  $<$1.5                  &  $<$2.5                  &  $1.7_{-1.2}^{+2.1}$       &  $2.0_{-0.9 }^{+2.0}$    \\
  \noalign{\smallskip} 
$\Gamma$                      & 1.19$_{-0.03 }^{+ 0.09}$&  $0.86_{-0.11}^{+0.11}$  &  $0.89_{-0.09}^{+0.12}$   &  $1.20_{-0.38}^{+0.40}$    &  $1.07_{-0.08}^{+0.19}$  &  $1.48_{-0.17}^{+0.29}$  &  $1.86_{-0.22}^{+0.34}$    &  $2.17_{-0.19}^{+0.38}$  \\
  \noalign{\smallskip} 
$E_{\rm p}$$^{\mathrm{d}}$    & 5.3$_{-0.8 }^{+ 1.4}$   &   $8.28_{-2.22}^{+4.00}$ &   $4.91_{-0.99}^{+1.22}$  &   $3.00_{-1.47}^{+4.34}$   &   $2.21_{-0.44}^{+0.58}$ &   $1.18_{-0.33}^{+0.58}$ &   $0.42_{-0.13}^{+0.43}$   &   $-$                    \\
  \noalign{\smallskip} 
d.o.f.                        & 385                     &    184                   &    242                    &   68                       &    159                   &    95                    &   118                      &    124                   \\
  \noalign{\smallskip} 
$\chi^2_{\rm r}$              & 1.04                    &   0.97                   &   1.05                    &   1.01                     &   0.97                   &   1.22                   &   0.92                     &   1.03                   \\

  \noalign{\smallskip} 
\hline
  \noalign{\smallskip} 
\multicolumn{9}{c}{Band model} \\
  \noalign{\smallskip} 
\hline
  \noalign{\smallskip} 

$N_{\rm H}$$^{\mathrm{c}}$    & $<$0.4                    & $<$0.6                     & $<$0.7                     &   $4.2_{-2.1}^{+2.5}$     &   $<$1.0                & $<$0.6                    & $<$1.0                     & $<$1.7                    \\ 
  \noalign{\smallskip} 
$\alpha_{\rm Band}$$^{\mathrm{e}}$ & $-1.11_{-0.09 }^{+ 0.09}$ &   $-0.86_{-0.11}^{+0.11}$  &   $-0.78_{-0.13}^{+0.16}$  &   $-1.20$                 &   $-1.07$               &   $-1.00$                 &   $-1.17_{-0.17}^{+0.07}$  &   $-1.63_{-0.15}^{+0.11}$ \\ 
  \noalign{\smallskip} 
$\beta_{\rm Band}$$^{\mathrm{e}}$  & $<-1.8$                   &   $<-4.5$                  &  $<-1.7$                   &   $<-1.9$                 &   $<-2.47 $             &   $-2.62_{-0.20}^{+0.15}$ &   $-2.52$                  &   $-2.93$                 \\ 
  \noalign{\smallskip} 
$E_{\rm p}$$^{\mathrm{d}}$    & 4.2$_{-0.6 }^{+ 1.43}$    & $8.34_{-1.36}^{+1.47}$     & $4.04_{-1.25}^{+1.60}$     & $2.63_{-0.55}^{+0.75}$    & $2.17_{-0.20}^{+0.14}$  & $1.11_{-0.10}^{+0.10}$    & $<$0.91                    & $<$0.64                   \\ 
  \noalign{\smallskip} 
d.o.f.                        & 384                       &  183                       &  241                       &   68                      &  159                    &   95                      &  118                       &  124                      \\ 
  \noalign{\smallskip} 
$\chi^2_{\rm r}$              & 1.03                      &  0.98                      &  1.03                      &  1.00                     &  0.96                   &  1.14                     &  0.95                      &  1.03                     \\ 

  \noalign{\smallskip} 
\hline
  \noalign{\smallskip} 
\multicolumn{9}{c}{Power-law model plus blackbody}\\
  \noalign{\smallskip} 
\hline
  \noalign{\smallskip} 

$N_{\rm H}$$^{\mathrm{c}}$    & $<$1.0                   &   $<$0.9                     &   $<$1.6                    &   $<$7.4                     &   $<$3.5                     &   $<$2.0                     &   $2.9_{-1.7}^{+1.7}$         &   $2.4_{-1.4}^{+1.5}$      \\       
  \noalign{\smallskip} 
$\Gamma$                      & 0.55$_{-0.06 }^{+ 0.08}$ &   $1.23_{-0.07}^{+0.09}$     &   $1.34_{-0.05}^{+0.09}$    &   $1.62_{-0.26}^{+0.23}$     &   $1.90_{-0.14}^{+0.13}$     &   $2.05_{-0.06}^{+0.15}$     &   $2.44_{-0.15}^{+0.14}$      &   $2.77_{-0.13}^{+0.13}$   \\               
  \noalign{\smallskip} 
$kT_{\rm bb}$$^{\mathrm{f}}$  & 1.94$_{-0.49 }^{+ 0.49}$ &   $0.89_{ -0.15}^{+ 0.16} $  &   $0.60_{ -0.06}^{+ 0.07} $ &   $0.465_{-0.069}^{+0.120}$  &   $0.47_{ -0.05}^{+ 0.06} $  &   $0.31_{ -0.03}^{+ 0.036}$  &   $0.35_{ -0.07}^{+ 0.08} $   &   $0.25_{ -0.03}^{+ 0.04}$ \\ 
  \noalign{\smallskip} 
$R_{\rm bb}$$^{\mathrm{g}}$  &  1.50$_{-0.05 }^{+ 0.06}$ &   $2.1_{-0.5}^{+0.7}$        &   $3.4_{-0.7}^{+0.6}$       &   $4.2_{-1.9}^{+1.9}$        &   $3.5_{-0.8}^{+1.0}$        &   $5.7_{-1.4}^{+1.1}$        &   $2.5_{-1.1}^{+1.6}$         &   $3.5_{-1.2}^{+1.4}$      \\               
  \noalign{\smallskip} 
d.o.f.                        & 384                      & 183                          & 241                         &  67                          & 158                          &  94                          & 117                           & 123                        \\               
  \noalign{\smallskip} 
$\chi^2_{\rm r}$              & 1.05                     &  0.95                        &  1.04                       &  0.96                        &  1.01                        &  1.19                        &  0.91                         &  1.00                      \\               
  \noalign{\smallskip} 
$F_{\rm bb}/F$$^{\mathrm{h}}$ & 0.10                     &  0.16                        &  0.17                       &  0.18                        &  0.22                        &  0.23                        &  0.10                         &  0.12                      \\               
  \noalign{\smallskip} 
Ftest$^{\mathrm{i}}$          & 1.9$\times10^{-11}$      &  2.4$\times10^{-7}$         &   7.9$\times10^{-10}$       &   1.8$\times10^{-20}$         &   1.5$\times10^{-10}$        &   1.0$\times10^{-5}$         &   3.3$\times10^{-4}$          &   3.1$\times10^{-5}$       \\              

  \noalign{\smallskip} 
  \hline
  \end{tabular}
  \end{center}
  \begin{list}{}{} 
  \item[$^{\mathrm{a}}$] Start and stop times of the time interval in seconds.
  \item[$^{\mathrm{b}}$] Width in pixels of the central box (20 pixel thick) that we excluded 
                         from the $40\times20$ pixel extraction region to account for the pile-up effect.
  \item[$^{\mathrm{c}}$] Extragalactic absorption column in units of $10^{20}$~cm$^{-2}$. The model included
                         both a {\tt wabs} component accounting for Galactic absorption (i.e. with $N_{\rm H}$ 
                         fixed to $3\times10^{20}$~cm$^{-2}$; \citealp{Dickey1990}) and a {\tt zwabs} component
                         with redshift fixed to $z=0.125$ and free $N_{\rm H}$ parameter to account for
                         extragalactic absorption.
  \item[$^{\mathrm{d}}$] Peak energy in units of keV. It is related to the cut-off energy $E_{\rm cut}$ through the equation
                         $E_{\rm p}=(2-\Gamma)E_{\rm cut}$ for the cut-off power law model (provided that $\Gamma < 2$) 
                         and $E_{\rm p}=(2+\alpha_{\rm Band})E_{\rm cut}$ for the Band model.
  \item[$^{\mathrm{e}}$] $\alpha_{\rm Band}$ and $\beta_{\rm Band}$ are the low and high energy indices of the Band model, respectively.
  \item[$^{\mathrm{f}}$] Blackbody temperature in units of keV.
  \item[$^{\mathrm{g}}$] Blackbody radius in units of $10^{11}$~cm. 
  \item[$^{\mathrm{h}}$] Unabsorbed flux of the blackbody component in the 0.2$-$10 keV range ($F_{\rm bb}$) over the total
                         unabsorbed flux ($F$). 
  \item[$^{\mathrm{i}}$] Chance probability of improvement of the fit adding a blackbody component to the absorbed power-law.
  \item Notes:
  \item {\it i)} The first column of the table corresponds to the integrated WT spectrum.
  \item {\it ii)} Parameter values are reported without errors when frozen in the fitting procedure. 
  \end{list} 
  \end{table*} 



\Online 

\setcounter{table}{0}
 \begin{table*}         
 \begin{center}         
 \caption{BAT and XRT data log.}     
 \label{log}
 \begin{tabular}{llllrr} 
 \hline 
 \hline 
 \noalign{\smallskip} 
  Sequence      &      Obs/Mode &         Start time  (UT)       &     End time   (UT)          & Exposure (s)  &  Start time$^{\mathrm{a}}$  \\
 \noalign{\smallskip} 
 \hline 
 \noalign{\smallskip} 

00214805000     &       BAT     &       2006-06-14 12:39:49     &       2006-06-14 12:51:51     &       759     &      -240     \\

 \noalign{\smallskip} 
 \hline          
 \noalign{\smallskip} 

00214805000     &       XRT/WT  &       2006-06-14 12:45:25     &       2006-06-14 20:23:16     &       414     &       97      \\
00214805000     &       XRT/PC  &       2006-06-14 13:57:39     &       2006-06-14 20:45:06     &       10628   &       4449    \\
00214805001     &       XRT/PC  &       2006-06-15 00:05:06     &       2006-06-15 23:59:57     &       17642   &       40878   \\
00214805002     &       XRT/PC  &       2006-06-16 00:08:04     &       2006-06-17 22:49:57     &       42687   &       127456  \\
00214805003     &       XRT/PC  &       2006-06-18 00:05:06     &       2006-06-20 23:06:58     &       57900   &       300078  \\
00214805004     &       XRT/PC  &       2006-06-21 00:30:32     &       2006-06-21 23:13:57     &       16781   &       560804  \\
00214805005     &       XRT/PC  &       2006-06-22 00:27:31     &       2006-06-22 23:25:58     &       20112   &       647023  \\
00214805006--08 &       XRT/PC  &       2006-06-23 02:10:33     &       2006-06-25 23:34:58     &       37183   &       739604  \\
00214805009--10 &       XRT/PC  &       2006-06-26 00:44:32     &       2006-06-27 23:34:56     &      28171    &       993644  \\
00214805011     &       XRT/PC  &       2006-06-27 12:21:21     &       2006-06-27 23:52:56     &       7356    &       1121852 \\
00214805012--17 &       XRT/PC  &       2006-06-28 01:03:23     &       2006-06-30 22:37:58     &       61535   &       1167575 \\
00214805018--19 &       XRT/PC  &       2006-07-01 01:35:19     &       2006-07-03 22:54:56     &       65247   &       1428691 \\
00214805020--21 &       XRT/PC  &       2006-07-04 00:12:31     &       2006-07-05 23:06:57     &       40088   &       1682922 \\
00214805022--23 &       XRT/PC  &       2006-07-06 00:27:57     &       2006-07-06 23:11:57     &       18101   &       1856649 \\
00214805024--28 &       XRT/PC  &       2006-07-07 00:31:50     &       2006-07-10 18:33:58     &       32353   &       1943282 \\
00214805029--30 &       XRT/PC  &       2006-07-14 05:38:50     &       2006-07-17 22:32:56     &       18815   &       2566501 \\
00214805031--37 &       XRT/PC  &       2006-07-19 04:51:31     &       2006-08-07 21:37:56     &       38602   &       2995662 \\
  \noalign{\smallskip}
  \hline
  \end{tabular}
  \end{center}
  \begin{list}{}{} 
  \item[$^{\mathrm{a}}$] Seconds since the trigger.
  \end{list} 
  \end{table*} 

\setcounter{table}{1}
 \begin{table*}         
 \begin{center}         
 \caption{UVOT data.}     
 \label{loguvot}
 \tiny
 \begin{tabular}{llllrrrr} 
 \hline 
 \hline 
 \noalign{\smallskip} 
 Sequence       &    Obs/Filter  &    Start time  (UT)        &    End time   (UT)   & Exposure (s)  &   Start time$^{\mathrm{a}}$ &  mag$^{\mathrm{b}}$   &  err    \\  
 \noalign{\smallskip} 
 \hline 
 \noalign{\smallskip}

 00214805000    &     UVOT/White  &    2006-06-14   12:45:33    &  2006-06-14   12:47:11    &           97      &      104    &     18.41  &  0.11    \\  
 00214805000    &       UVOT/$V$  &    2006-06-14   12:47:15    &  2006-06-14   12:50:30    &           191     &     206.5   &  $>$19.70  &   --     \\  
 00214805000    &      UVOT/UVM2  &    2006-06-14   13:57:40    &  2006-06-14   14:00:60    &            22     &     4431    &  $>$17.70  &   --     \\  
 00214805000    &      UVOT/UVW1  &    2006-06-14   14:01:05    &  2006-06-14   14:04:25    &           197     &     4636    &     18.49  &  0.31    \\  
 00214805000    &       UVOT/$U$  &    2006-06-14   14:04:29    &  2006-06-14   14:07:49    &           197     &     4840    &     18.79  &  0.16    \\  
 00214805000    &       UVOT/$B$  &    2006-06-14   14:07:54    &  2006-06-14   14:11:14    &           197     &     5045    &     19.82  &  0.21    \\  
 00214805000    &     UVOT/White  &    2006-06-14   14:11:18    &  2006-06-14   14:14:38    &           197     &     5249    &     18.25  &  0.06    \\  
 00214805000    &      UVOT/UVW2  &    2006-06-14   14:14:43    &  2006-06-14   14:18:03    &           197     &     5454    &     18.26  &  0.23    \\  
 00214805000    &       UVOT/$V$  &    2006-06-14   14:18:08    &  2006-06-14   14:21:28    &           197     &     5659    &     19.34  &  0.27    \\  
 00214805000    &      UVOT/UVM2  &    2006-06-14   14:21:32    &  2006-06-14   14:24:52    &           197     &     5863    &     17.78  &  0.18    \\  
 00214805000    &      UVOT/UVW1  &    2006-06-14   14:24:57    &  2006-06-14   14:28:17    &           197     &     6068    &     18.32  &  0.25    \\  
 00214805000    &       UVOT/$U$  &    2006-06-14   14:28:22    &  2006-06-14   14:30:57    &           153     &     6272.5  &     18.55  &  0.19    \\  
 00214805000    &       UVOT/$B$  &    2006-06-14   15:34:07    &  2006-06-14   15:39:07    &           295     &     10218   &     19.46  &  0.13    \\  
 00214805000    &       UVOT/$B$  &    2006-06-14   15:39:11    &  2006-06-14   15:44:11    &           295     &     10522   &     19.51  &  0.13    \\  
 00214805000    &       UVOT/$B$  &    2006-06-14   15:44:13    &  2006-06-14   15:49:13    &           295     &     10824   &     19.56  &  0.12    \\  
 00214805000    &     UVOT/White  &    2006-06-14   15:49:18    &  2006-06-14   15:54:18    &           293     &     11129   &     18.34  &  0.07    \\  
 00214805000    &     UVOT/White  &    2006-06-14   15:54:21    &  2006-06-14   15:59:21    &           295     &     11432   &     18.36  &  0.05    \\  
 00214805000    &     UVOT/White  &    2006-06-14   15:59:25    &  2006-06-14   16:04:25    &           295     &     11736   &     18.24  &  0.08    \\  
 00214805000    &      UVOT/UVW2  &    2006-06-14   16:04:30    &  2006-06-14   16:13:26    &           528     &     12041   &     18.10  &  0.48    \\  
 00214805000    &       UVOT/$V$  &    2006-06-14   17:10:30    &  2006-06-14   17:15:30    &           295     &     16001   &     19.46  &  0.25    \\  
 00214805000    &       UVOT/$V$  &    2006-06-14   17:15:34    &  2006-06-14   17:20:34    &           295     &     16305   &     19.21  &  0.21    \\  
 00214805000    &       UVOT/$V$  &    2006-06-14   17:20:37    &  2006-06-14   17:25:37    &           295     &     16608   &     19.32  &  0.22    \\  
 00214805000    &      UVOT/UVM2  &    2006-06-14   17:25:41    &  2006-06-14   17:40:41    &           886     &     16912   &     18.62  &  0.81    \\  
 00214805000    &      UVOT/UVW1  &    2006-06-14   17:40:48    &  2006-06-14   17:52:21    &           682     &     17819.5 &     18.44  &  0.27    \\  
 00214805000    &       UVOT/$U$  &    2006-06-14   18:46:57    &  2006-06-14   18:51:57    &           295     &     21788   &     18.56  &  0.12    \\  
 00214805000    &       UVOT/$U$  &    2006-06-14   18:51:60    &  2006-06-14   18:56:60    &           295     &     22091   &     18.48  &  0.12    \\  
 00214805000    &       UVOT/$U$  &    2006-06-14   18:57:03    &  2006-06-14   19:02:03    &           295     &     22394   &     18.53  &  0.11    \\  
 00214805000    &       UVOT/$B$  &    2006-06-14   19:02:09    &  2006-06-14   19:07:09    &           295     &     22700   &     19.54  &  0.14    \\  
 00214805000    &       UVOT/$B$  &    2006-06-14   19:07:11    &  2006-06-14   19:12:11    &           295     &     23002   &     19.37  &  0.13    \\  
 00214805000    &       UVOT/$B$  &    2006-06-14   19:12:15    &  2006-06-14   19:17:15    &           295     &     23306   &     19.50  &  0.16    \\  
 00214805000    &     UVOT/White  &    2006-06-14   19:17:19    &  2006-06-14   19:22:19    &           295     &     23610   &     18.25  &  0.13    \\  
 00214805000    &     UVOT/White  &    2006-06-14   19:22:23    &  2006-06-14   19:27:23    &           295     &     23914   &     18.26  &  0.11    \\  
 00214805000    &     UVOT/White  &    2006-06-14   19:27:25    &  2006-06-14   19:28:46    &            80     &     24216.5 &     18.31  &  0.22    \\  
 00214805000    &      UVOT/UVW1  &    2006-06-14   20:23:21    &  2006-06-14   20:38:21    &           886     &     27572   &     18.38  &  0.12    \\  
 00214805000    &       UVOT/$U$  &    2006-06-14   20:38:27    &  2006-06-14   20:43:27    &           295     &     28478   &     18.34  &  0.10    \\  
 00214805000    &       UVOT/$U$  &    2006-06-14   20:43:31    &  2006-06-14   20:45:11    &            99     &     28782   &     18.45  &  0.21    \\  
 00214805001    &      UVOT/UVW1  &    2006-06-15   00:01:34    &  2006-06-15   02:54:47    &           396     &     40665.5 &     19.02  &  0.26    \\  
 00214805001    &       UVOT/$U$  &    2006-06-15   00:03:54    &  2006-06-15   02:55:59    &           197     &     40805.5 &     18.76  &  0.17    \\  
 00214805001    &       UVOT/$B$  &    2006-06-15   00:05:05    &  2006-06-15   02:57:10    &           197     &     40876.5 &     19.73  &  0.19    \\  
 00214805001    &      UVOT/UVW2  &    2006-06-15   00:06:19    &  2006-06-15   03:01:45    &           794     &     40950   &     18.74  &  0.46    \\  
 00214805001    &       UVOT/$V$  &    2006-06-15   00:10:53    &  2006-06-15   03:02:57    &           197     &     41224   &     19.23  &  0.28    \\  
 00214805001    &      UVOT/UVM2  &    2006-06-15   00:12:04    &  2006-06-15   03:06:01    &           527     &     41295.5 &     18.65  &  0.54    \\  
 00214805001    &      UVOT/UVW1  &    2006-06-15   04:29:34    &  2006-06-15   07:44:28    &           375     &     56745   &     19.17  &  0.26    \\  
 00214805001    &       UVOT/$U$  &    2006-06-15   04:31:52    &  2006-06-15   07:45:29    &           187     &     56883.5 &     19.29  &  0.21    \\  
 00214805001    &       UVOT/$B$  &    2006-06-15   04:33:03    &  2006-06-15   07:46:31    &           187     &     56954   &     19.90  &  0.22    \\  
 00214805001    &       UVOT/VW2  &    2006-06-15   04:34:16    &  2006-06-15   07:50:26    &           753     &     57027   &     19.63  &  0.27    \\  
 00214805001    &       UVOT/$V$  &    2006-06-15   04:38:49    &  2006-06-15   07:51:28    &           187     &     57300.5 &  $>$19.80  &   --     \\  
 00214805001    &      UVOT/UVM2  &    2006-06-15   04:40:02    &  2006-06-15   07:54:02    &           502     &     57373   &     18.75  &  0.33    \\  
 00214805001    &      UVOT/UVW1  &    2006-06-15   09:18:34    &  2006-06-15   12:34:28    &           474     &     74085   &     19.35  &  0.22    \\  
 00214805001    &       UVOT/$U$  &    2006-06-15   09:20:52    &  2006-06-15   12:35:59    &           237     &     74223.5 &     19.33  &  0.19    \\  
 00214805001    &       UVOT/$B$  &    2006-06-15   09:22:04    &  2006-06-15   12:37:32    &           237     &     74295   &     20.35  &  0.27    \\  
 00214805001    &      UVOT/UVW2  &    2006-06-15   09:23:15    &  2006-06-15   12:43:26    &           950     &     74366.5 &     19.75  &  0.60    \\  
 00214805001    &       UVOT/$V$  &    2006-06-15   09:27:49    &  2006-06-15   12:44:59    &           237     &     74640   &     19.77  &  0.35    \\  
 00214805001    &      UVOT/UVM2  &    2006-06-15   09:28:60    &  2006-06-15   12:49:02    &           648     &     74711   &     19.04  &  0.37    \\  
 00214805001    &      UVOT/UVW1  &    2006-06-15   14:07:33    &  2006-06-15   17:24:57    &           769     &     91424   &     19.68  &  0.23    \\  
 00214805001    &       UVOT/$U$  &    2006-06-15   14:11:51    &  2006-06-15   17:27:15    &           384     &     91682.5 &     19.61  &  0.18    \\  
 00214805001    &       UVOT/$B$  &    2006-06-15   14:14:04    &  2006-06-15   17:29:32    &           384     &     91815   &     20.48  &  0.23    \\  
 00214805001    &      UVOT/UVW2  &    2006-06-15   14:16:18    &  2006-06-15   17:38:25    &          1540     &     91949.5 &     20.52  &  0.43    \\  
 00214805001    &       UVOT/$V$  &    2006-06-15   14:24:52    &  2006-06-15   17:40:44    &           384     &     92463   &  $>$20.30  &   --     \\  
 00214805001    &      UVOT/UVM2  &    2006-06-15   14:27:03    &  2006-06-15   17:47:03    &          1086     &     92594.5 &     19.78  &  0.60    \\  
 00214805001    &      UVOT/UVW1  &    2006-06-15   18:56:33    &  2006-06-15   22:13:57    &           779     &     108764  &     20.45  &  0.32    \\  
 00214805001    &       UVOT/$U$  &    2006-06-15   19:01:03    &  2006-06-15   22:16:13    &           389     &     109034  &     20.72  &  0.35    \\  
 00214805001    &       UVOT/$B$  &    2006-06-15   19:03:19    &  2006-06-15   22:18:31    &           387     &     109170  &     20.77  &  0.27    \\  
 00214805001    &      UVOT/UVW2  &    2006-06-15   19:05:37    &  2006-06-15   22:27:27    &          1560     &     109308  &     20.50  &  0.40    \\  
 00214805001    &       UVOT/$V$  &    2006-06-15   19:14:33    &  2006-06-15   22:29:45    &           389     &     109844  &  $>$20.30  &   --     \\  
 00214805001    &      UVOT/UVM2  &    2006-06-15   19:16:49    &  2006-06-15   22:36:01    &          1101     &     109980  &     20.47  &  0.74    \\  
 00214805002     &     UVOT/UVW2   &   2006-06-16   01:32:19    & 2006-06-17   21:03:49     &         9481      &    132510   &     21.25  &  0.19    \\  
 00214805002     &     UVOT/UVW1   &   2006-06-16   00:08:07    & 2006-06-17   22:28:05     &         7050      &    127458   &     21.24  &  0.16    \\  
 00214805002     &      UVOT/$U$   &   2006-06-16   00:10:31    & 2006-06-17   22:30:21     &         3516      &    127602   &     21.66  &  0.26    \\  
 00214805002     &      UVOT/$B$   &   2006-06-16   00:11:44    & 2006-06-17   22:32:38     &         3520      &    127675   &  $>$22.50  &   --     \\  
 00214805002     &      UVOT/$V$   &   2006-06-16   00:17:41    & 2006-06-17   22:43:51     &         3520      &    128032   &     21.42  &  0.34    \\  
 00214805002     &     UVOT/UVM2   &   2006-06-16   00:18:54    & 2006-06-17   22:50:02     &         9880      &    128105   &     21.51  &  0.21    \\  
 00214805003     &     UVOT/UVW2   &   2006-06-18   00:13:06    & 2006-06-20   21:20:60     &        13506      &    300557   &     22.74  &  0.31    \\  
 00214805003     &     UVOT/UVW1   &   2006-06-18   00:03:41    & 2006-06-20   22:44:55     &        10151      &    299992   &     22.71  &  0.36    \\  
 00214805003     &      UVOT/$U$   &   2006-06-18   00:08:19    & 2006-06-20   22:47:13     &         4937      &    300270   &     22.20  &  0.34    \\  
 00214805003     &      UVOT/$B$   &   2006-06-18   00:10:42    & 2006-06-20   22:49:30     &         4741      &    300413   &  $>$22.70  &   --     \\  
 00214805003     &      UVOT/$V$   &   2006-06-18   00:22:20    & 2006-06-20   23:00:48     &         4674      &    301111   &  $>$21.70  &   --     \\  
 00214805003     &     UVOT/UVM2   &   2006-06-18   00:24:42    & 2006-06-20   23:07:02     &        12910      &    301253   &     22.73  &  0.33    \\  
 00214805007--09 &      UVOT/$V$   &   2006-06-24   15:27:35    & 2006-06-26   23:18:25     &        11028      &    873826   &  $>$22.10  &   --     \\  
 00214805008--09 &      UVOT/$B$   &   2006-06-25   00:50:17    & 2006-06-26   23:30:02     &         6959      &   907588.5  &  $>$23.10  &   --     \\  
 00214805011--27 &     UVOT/White  &   2006-06-27   12:21:24    & 2006-07-10   23:26:12     &         9012      &   1121855   &     22.38  &  0.22    \\  
 00214805008--35 &      UVOT/$U$   &   2006-06-25   00:44:08    & 2006-08-02   00:00:02     &       184470      &    907219   &     23.70  &  0.23    \\  
 00214805008--37 &     UVOT/UVW1   &   2006-06-25   00:56:30    & 2006-08-07   16:49:01     &        17548      &   907961.5  &  $>$23.10  &   --     \\  
 00214805036--37 &     UVOT/UVM2   &   2006-08-03   00:08:56    & 2006-08-07   16:47:48     &         5773      &   4274707   &  $>$22.50  &   --     \\  
 00214805036--37 &     UVOT/UVW2   &   2006-08-03   01:58:13    & 2006-08-07   21:38:01     &         3802      &   4281264   &  $>$22.50  &   --     \\  

  \noalign{\smallskip}
  \hline
  \end{tabular}
  \end{center}
  \begin{list}{}{} 
  \item[$^{\mathrm{a}}$] Seconds since the trigger.
  \item[$^{\mathrm{b}}$] Magnitude corrected for Galactic extinction.
  \end{list} 
  \end{table*} 

\end{document}